\documentclass[10pt,twocolumn,twoside]{IEEEtran}

\usepackage{array}
\newcolumntype{P}[1]{>{\centering\arraybackslash}p{#1}}
\usepackage{float}
\usepackage{tabu}
\usepackage{amsmath}

\usepackage{graphicx}
\usepackage{blindtext}
\usepackage{epsfig} 
\usepackage{mathptmx} 
\usepackage{times} 
\usepackage{amssymb}  
\usepackage{multirow}
\usepackage{subcaption}
\usepackage{csquotes}
\usepackage{algpseudocode}
\usepackage{algorithm}
\usepackage{tikz}
\usetikzlibrary{shapes.geometric, arrows}
\usepackage{longtable}
\usepackage{amsbsy}
\usepackage{hyperref}
\usepackage{authblk}
\usepackage{leftidx}
\usepackage{ragged2e}
\usepackage[font={scriptsize}]{caption}
\usepackage{color}

\begin{document}

\title{A Satisficing Control Design Framework with Safety and Performance Guarantees for Constrained Systems under Disturbances}

%
\markboth{ }
{Han \MakeLowercase{\textit{et al.}}: A Satisficing Control Design Framework with Safety and Performance Guarantees for Constrained Systems under Disturbances}
\author{Yuzhen Han, \textit{Student Member, IEEE} and Hamidreza Modares, \textit{Senior Member, IEEE} 
\thanks{
Yuzhen Han and Hamidreza Modares are with  the Department
of Mechanical Engineering, Michigan State University, East Lansing, MI, 48863, USA (e-mails:hanyuzh1@msu.edu; modaresh@msu.edu).}
}

\maketitle

\begin{abstract}
This paper presents a safe robust policy iteration (SR-PI) algorithm to design controllers with satisficing (good enough) performance and safety guarantee. This is in contrast to standard PI-based control design methods with no safety certification. It also moves away from existing safe control design approaches that perform pointwise optimization and are thus myopic. Safety assurance requires satisfying a control barrier function (CBF), which might be in conflict with the performance-driven Lyapunov solution to the Bellman equation arising in each iteration of the PI. Therefore, a new development is required to robustly certify the safety of an improved policy at each iteration of the PI. The proposed SR-PI algorithm unifies performance guarantee (provided by a Bellman inequality) with safety guarantee (provided by a robust CBF) at each iteration. The Bellman inequality resembles the satisficing decision making framework and parameterizes the sacrifice on the performance with an aspiration level when there is a conflict with safety. This aspiration level is optimized at each iteration to minimize the sacrifice on the performance. It is shown that the presented satisficing control policies obtained at each iteration of the SR-PI guarantees robust safety and performance. Robust stability is also guaranteed when there is no conflict with safety. Sum of squares (SOS) program is employed to implement the proposed SR-PI algorithm iteratively. Finally, numerical simulations are carried out to illustrate the proposed satisficing control framework.
\end{abstract}
\vspace{-0.1cm}


\IEEEpeerreviewmaketitle
\vspace{-0.25cm}
\section{Introduction}

\IEEEPARstart{S}{uccessful} deployment of the next-generation safety-critical systems (e.g., self-driving cars or assistive robots) requires certifying their safety despite uncertainties \cite{a1,a2}. Therefore, there is an urgent need for developing robust safe controllers that respect the system’s constraints all the time. Safety, however, is the bare minimum requirement for any safety-critical system, and it is desired to design safe controllers that achieve as much performance as possible. To guarantee performance, one can solve an optimal control problem for which a pre-defined cost function that encodes desired system specifications is optimized \cite{smc1,smc2}. Despite the importance of designing safe optimal controllers, safe control design and optimal control design are typically separated in the literature. More specifically, while an optimal controller is generally found by solving the so-called Hamilton-Jacobi-Bellman (HJB) equation \cite{a3,a4}, existing iterative solutions to the HJB equation mainly ignore safety constraints. On the other hand, to satisfy safety requirement of dynamic systems, safety verification using control barrier functions (CBFs) \cite{a5,a6,a7} and reachability analysis \cite{a8,a9,a10} have been widely and successfully used. While these frameworks can effectively guarantee the forward invariance of a given safe set and asymptotic convergence of the system’s trajectories to a target set, the long-term optimality of the solution is not considered. This  can lead to conservative solutions that unnecessarily consume a significant amount of resources or result in poor performance.

To take into account safety and optimality simultaneously, in the reference governor approach \cite{a11,a12}, a safe controller intervenes with a nominal performance-driven controller to avoid constraint violations when there is a risk of safety violation. However, reference governor might keep intervening with the nominal controller, making the controller myopic and possibly far from optimal. This is because the nominal controller does not take into account the safety constraints in the design phase to proactively avoid them. Model predictive control (MPC) \cite{a13,a14} is another control strategy candidate for addressing safe optimal control design. MPC takes the system constraints into account while optimizing a short-horizon performance. However, despite its tremendous success, MPC needs to perform an optimization algorithm at every step, which might not be computationally tractable for nonlinear systems. Moreover, because of its short-sided nature, guaranteeing feasibility and stability is also hard.  

Satisficing decision theory \cite{a15} has been widely employed in economic optimization problems to find good enough solutions that are not necessarily optimal. This is motivated by the fact that finding optimal solutions for systems under limited resources and incomplete information might not be feasible. Satisficing stabilizing control design has also been considered in control community \cite{a16,a17}. These approaches, however, are typically plagued by the lack of safety guarantee. In this paper, we propose a new safe and satisficing control approach in which a satisficing framework is leveraged to sacrifice performance in favor of safety, but minimizing the sacrifice level on the performance as much as possible. Starting from a safe control policy, an iterative safe robust policy iteration (SR-PI) algorithm is then proposed to find improved satisficing controllers that certify robust safety against matched disturbances. More specifically, the policy evaluation step finds the value function for the current the satisficing safe control policy and the policy improvement step finds an improved satisficing control policy with guaranteed input-to-state safety (ISSf) \cite{a18}. The robust stability of satisficing policies will also be guaranteed when there is no conflict between safety and stability. Sum of squares (SOS) program \cite{a19} is employed to implement the presented PI algorithm. Fig. 1 shows the schematic of the proposed SR-PI and its comparison with the standard PI algorithm. It should be noticed that the type matched disturbance is investigated in this research, but it could be generalized to unmatched disturbance case which also attracts wide attentions\cite{unmatch1,unmatch2,unmatch3}.

The rest of the paper is organized as follows. Section 2 presents some preliminaries that are used throughout the paper. A satisficing safe control design framework is developed in Section 3. Sections 4 presents the simulation results and experiment comparison, respectively. The paper is concluded in Section 5. 

\textit{Notations:} Through the paper, the set of continuously differentiable functions are represented by  ${{C}^{1}}$, and the set of positive definite and proper functions in ${{C}^{1}}$ are denoted as $P$. A polynomial $p(x)$ is a sum of squares (SOS) polynomial, i.e., $p(x)\in {{\mathcal{P}}^{SOS}}$ where ${{\mathcal{P}}^{SOS}}$ is a set of SOS polynomial, if $p(x)=\sum\nolimits_{1}^{m}{p_{i}^{2}(x)}$ where $p_{i}^{{}}(x)\in {{\mathcal{P}}^{SOS}}$, $i=1,...,m$. $\mathbb{R}{{[x]}_{{{d}_{1}},{{d}_{2}}}}$ denotes all the sets of polynomials in $x\in {{\mathbb{R}}^{n}}$ with degree at least ${{d}_{1}}$ and at most ${{d}_{2}}$. $\mathcal{X}\in {{\mathbb{R}}^{n}}$ is state space which is a compact set. A continuous function $K:[0,a)\to [0,\infty )$ is of class $\kappa $ function, denoted by $K\in \kappa $, if it is strictly increasing and $K(0)=0$. A function $\alpha (s,t)$ is a class of $\kappa \mathcal{L}$ function if for each fixed $t\ge 0$ the function $\alpha (.,t)$ is a $\kappa$ function and for fixed  $s\ge 0$ it is decreasing to zero as $t\to \infty $. We also denote by $\kappa \kappa$ all functions $\gamma $ such that $\gamma (.,t)\in \kappa $ for a fixed $t\ge 0$ and similarly, $\gamma (s,.)\in \kappa $ for a fixed $s\ge 0$. $\nabla f(x)$ is the gradient of function $f$ and  $\nabla f(x)={{[\frac{\partial f(x)}{\partial {{x}_{1}}},\frac{\partial f(x)}{\partial {{x}_{1}}},...,\frac{\partial f(x)}{\partial {{x}_{n}}}]}^{T}}$. $diag({{x}_{1}},...,{{x}_{n}})$ denotes a square diagonal matrix with elements of ${{x}_{1}},...,{{x}_{n}}$ on the main diagonal. $||x||$ indicates the Euclidean norm $\sqrt{{{x}^{T}}x}$ of a real vector $x\in {{\mathbb{R}}^{n}}$. For any set $S$, $Int(S)$ and $\partial S$ denote the interior and boundary of the set $S$, respectively; $\overline{Int(S)}$ is the closure of set $S$. $||\xi |{{|}_{U}}={{\min }_{a\in U}}||\xi -a||$ where $||.||$ is Euclidian norm. For a given signal $x:\mathbb{R}\to {{\mathbb{R}}^{n}}$, its ${{L}^{P}}$ norm on the interval $T$ is given by $||x|{{|}_{{{L}^{P}}(T)}}={{({{\int_{T}{||x(t)||}}^{P}}dt)}^{1/P}}$ and similarly, its ${{L}^{\infty }}$ norm is defined by $||x|{{|}_{{{L}^{\infty }}(T)}}=(ess){{\sup }_{t\in T}}(||x(t)||)$. For the sake of conciseness, for $T=[0,\infty )$, we denote the ${{L}^{\infty }}$ norm of $x$ simply by $||x|{{|}_{{{L}^{\infty }}}}$. For two vectors $x$ and $y$, $x\succeq y$ iff $x_i \geq y_i$ holds for all elements $x_i$ and $y_i$ of $x$ and $y$. $f_{max}$ indicates the maximum of the function $f$ over a set of interest.

\vspace{-0.2cm}
\begin{figure}[!ht]
\begin{center}
\includegraphics[width=0.65\columnwidth,height=65mm]{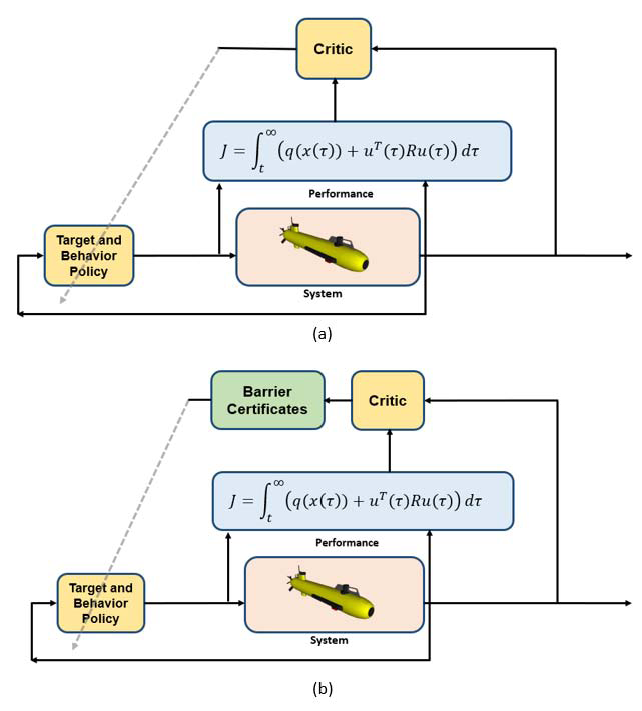}
\vspace{-5pt}\caption{Comparison between the proposed SR-PI algorithm and existing PI algorithms. (a) Standard PI without safety verification; (b) The proposed SR-PI with safety verification at each iteration to find an improved safe policy.}
\label{illu}
\captionsetup{justification=centering}
\end{center}
\vspace{-0.2cm}
\end{figure}

\vspace{-0.1cm}
\section{Preliminaries}
\vspace{-0.1cm}

Consider the following continuous-time nonlinear system
\begin{equation}
\dot{x}=f\left( x \right)+g\left( x \right)(u+d )=f\left( x \right)+g\left( x \right)u+\omega,\,\,\,\ 
\label{eq1}
\end{equation}

\vspace{-0.15cm}
\noindent
where $x\in \mathcal{X}$ is the vector of system states, $u\in {{\mathbb{R}}^{\text{m}}}$  is the vector of control input, $\omega $ is the disturbance on the control input, and $\omega=g\left( x \right)d $ . The nonlinear functions $f:{{\mathbb{R}}^{\text{n}}}\to {{\mathbb{R}}^{\text{n}}}$ and $\text{g}:{{\mathbb{R}}^{\text{n}}}\to {{\mathbb{R}}^{\text{n}\times \text{m}}}$ are assumed to be locally Lipschitz continuous with $f\left( 0 \right)=0$.
\smallskip

\noindent
$\textbf{Assumption 1}$ The system (\ref{eq1}) is stabilizable on the set $\mathcal{X}$.  \vspace{2pt}

\smallskip

\noindent


\smallskip

\noindent
\textbf{Assumption 2} The disturbance $d$ is bounded. That is, there exists a constant $d_{\max }^{{}}$ such that

\vspace{-0.2cm}
\begin{equation}
||d(t)||\le d_{\max }^{{}}
\label{eq3}
\end{equation}

\vspace{-0.15cm}

\vspace{-0.15cm}

\subsection{Optimal Control Design Framework}
In this section, we present a robust optimal control design framework for the system (\ref{eq1}). To find an optimal controller, one can optimize a pre-defined performance index that encodes the designer’s intention in achieving some system’s specifications. For the case where $d=0$,  the following infinite-horizon performance index is usually considered for the system (\ref{eq1}).

\vspace{-0.3cm}
\begin{equation}
J\left( {x},u \right)=\mathop{\int }_{t}^{\infty }\,r\left( x,u \right)d\tau
\label{eq4}
\end{equation}

\vspace{-0.1cm}
\noindent
where

\vspace{-0.3cm}
\begin{equation}
r\left( x,u \right)=q\left( x \right)+{{u}^{T}}Ru\text{ }\!\!\!\!\text{ }
\label{eq5}
\end{equation}

\noindent
is the reward function with $q(x)\in \mathbb{R}$ as a positive definite function, and $R\in {{\mathbb{R}}^{m\times m}}$ as a symmetric positive definite matrix. The existence of a stabilizing optimal controller is guaranteed under some mild assumptions on the system dynamics and the performance index \cite{a20,smc3}. The optimal control found by optimizing (\ref{eq4}), however, does not guarantee robust stability of the system due to the disturbance. To guarantee robust stability of the system (\ref{eq1}) for the case when $d\ne 0$, the performance index  (\ref{eq4}) can be modified as follows \cite{a21}

\vspace{-0.1cm}
\begin{equation}
\overline{J}\left( {x},u \right)=\mathop{\int }_{t}^{\infty }\,\overline{r}\left( x,u \right)d\tau
\label{eq6}
\end{equation}

\vspace{-0.15cm}
\noindent
with the modified reward function

\begin{equation}
\overline{r}\left( x,u \right)=q\left( x \right)+{{u}^{T}}Ru+\beta \text{(x) }\!\!\!\!\text{ }
\label{eq7}
\end{equation}

\noindent
where $\beta \text{(x)}$ is an extra term added to guarantee robust stability despite the disturbance. \vspace{3pt}

\noindent
{\textbf{Remark 1.} Note that several modified performance or reward functions are presented in the presence of the disturbance. For example, the $H_\infty$ control defines the extra term as $\beta \text{(x)}=-\gamma^2 d(x)^T d(x)$. On the other hand, in \cite{a23}, the extra term is defined as $\beta \text{(x)=}\frac{1}{4}\nabla {{V}^{T}}\nabla V+d_{\max }^{2}$, where $V(x)$ is the value function corresponding to the control policy $u$, and it is shown that the optimal controller found by minimizing the modified cost guarantees robust stability and provides an upper bound for the original performance. However, since the extra term $\beta \text{(x)}$ depends on the gradient of the value of the control input quadratically, solving the modified optimal control problem becomes computationally expensive when SOS is used to implement it. In the following, a modified performance index is presented to avoid this issue. As will be shown later, instead of solving a huge SOS optimization (because of the cross term $\nabla {{V}^{T}}\nabla V$), two SOS optimizations with much less complexity will be solved. 

\smallskip
\noindent
\textbf{Theorem 1.} Consider the system (\ref{eq1}) and the performance function (\ref{eq6}) and (\ref{eq7}) and let

\vspace{-0.25cm}
\begin{equation}
\beta \text{(x)=}\beta_u(x)+d_{\max }^{T}Rd_{\max }
\label{eq9}
\end{equation}

\noindent with $\beta_u(x) \ge u_{\max }^{T}R \,\, {u}_{\max }$. Then, the optimal control solution is 
\begin{equation}
{{u}^{o}}(x)=-\frac{1}{2}{{R}^{-1}}{{({{(\nabla {{V}^{o}})}^{T}}\left( x \right)g(x))}^{T}}
\label{eq12}
\end{equation}
where $V^{o}$ is the solution to the HJB equation given by

\begin{equation}
\overline{H}\text{(}{{V}^{o}}\text{)=0}
\label{eq10}
\end{equation}

\noindent 
with 

\vspace{-0.3cm}
\begin{align}
\begin{gathered}
\overline{H}\text{(}V\text{)=}q(x)+\nabla {{V}^{T}}\left( x \right)f(x) \hfill \\
-\frac{1}{4}\nabla {{V}^{T}}\left( x \right)g(x){{R}^{-1}}{{(\nabla {{V}^{T}}\left( x \right)g(x))}^{T}}+d_{\max }^{T}Rd_{\max }+\beta_u(x) \hfill 
\end{gathered} 
\label{eq11}
\end{align}   

\noindent
That is, 

\vspace{-0.45cm}
\begin{equation}
{{V}^{o}}({{x}_{0}})=\underset{u}{\mathop{\min }}\,\overline{J}({{x}_{0}},u)=\overline{J}({{x}_{0}},{{u}^{o}})
\label{eq13}
\end{equation}

\noindent
 Moreover, the optimal controller is unique and guarantees robust stability and provides a suboptimal performance (i.e., an upper bound) for the original cost function (\ref{eq4}) and (\ref{eq5}).   \vspace{3pt}

\noindent
\textit{Proof.} The fact that the optimal control policy found by (\ref{eq12}) and (\ref{eq11}) can be shown similar to \cite{a20}, as only the cost function is modified and the derivation of the HJB equation does not change. We now show that the optimal controller guarantees robust stability and provided an upper bound for the original cost. First, we show that (\ref{eq12}) is a solution to the roust control problem. That is, the system (\ref{eq1}) is globally asymptotic stable for $u^o(x)$. To do this, we show that $V^{o}(x)$ is a Lyapunov function. Clearly,

\begin{equation}
\left\{ \begin{array}{l}
V^{o}(x)>0,\,\,\,\,\,x\ne 0 \\
V^{o}(x)=0,\,\,\,\,\,x=0 
\end{array}\right.
\label{proof1}
\end{equation}

\smallskip
\smallskip
\noindent
Also, $\dot{V}^{o}(x)=dV^{o}(x)/dt<0$ for $x \neq 0$, because

\vspace{-0.3cm}
\begin{align}
\begin{gathered}
    \dot{V}^{o}(x)={{(dV^{o}(x)/dx)}^{T}}(dx/dt) \hfill\\ 
  \quad \quad \quad ={{(\nabla V^{o})}^{T}}(x)(f(x)+g(x)u^{o}+w) \hfill\\
  \quad \quad \quad ={{(\nabla V^{o})}^{T}}(x)(f(x)+g(x)u^{o})+{{(\nabla V^{o})}^{T}}(x)g(x)d \hfill\\  
   \quad \quad \quad =-\beta _{u}^{{}}(x)-d_{\max }^{T}R{{d}_{\max }}-q(x)-{{u}^{{o}^{T}}}Ru^{o}+{{(\nabla V^{o})}^{T}}(x)g(x)d \hfill\\ 
  \quad \quad \quad =-\beta _{u}^{{}}(x)-d_{\max }^{T}R{{d}_{\max }}-q(x)-{{u}^{{o}^{T}}}Ru^{o}-2u^{o^T}Rd \hfill\\ 
  \quad \quad \quad =-\beta _{u}^{{}}(x)-d_{\max }^{T}R{{d}_{\max }} -q(x)-{{(d+u^{o}(x))}^{T}}R(d+u^{o}(x))\hfill\\ \quad \quad \quad \quad   +{{d}^{T}}Rd \hfill\\ 
  \quad \quad \quad \le -\beta _{u}^{{}}(x)-(d_{\max }^{T}R{{d}_{\max }}-{{d}^{T}}Rd)-q(x) \hfill\\
  \quad \quad \quad \le -\beta _{u}^{{}}(x)-q(x)<0. \hfill\\  
\end{gathered} 
\label{proof2}
\end{align}   

\noindent 
Therefore, the conditions of the Lyapunov local stability theory are satisfied. Consequently, there exists a constant $c > 0$ and a neighborhood $\mathcal{N}=\{x:||\beta_u(x)+q(x)||<c\}$ such that if $x(t)$ enters $\mathcal{N}$, then $lim_{t \to \infty} x(t)=0$. However, $x(t)$ cannot remain outside $\mathcal{N}$ forever. Otherwise, $||x(t)||\ge c$ for all $t \ge 0$. Therefore

\begin{align}
\begin{gathered}
   V(x(t))-V(x(0))=\int_{0}^{t}{\dot{V}(x(\tau ))}d\tau  \hfill\\  
  \quad \quad \quad\le -\int_{0}^{t}{||\beta_u(x)+q(x)|{{|}^{}}}d\tau  \hfill\\  
 \quad \quad \quad\le -\int_{0}^{t}{{{c}^{}}}d\tau ={{c}^{}}t \hfill\\  
 \end{gathered} 
\label{proof3}
\end{align}

Let $t \to +\infty$, we have 

\begin{align}
\begin{gathered}
   V(x(t))\leq V(x(0))-ct \to -\infty
 \end{gathered} 
\label{proof4}
\end{align}

 \noindent
which contradicts the fact that $V(x(t))\ge 0$ for all $x(t)$. Therefore $lim_{t \to \infty} x(t)=0$ no matter where the trajectory begins. So the optimal controller guarantees robust stability. From the (\ref{proof2}), the following holds

\begin{align}
\begin{gathered}
   \dot{V}(x)\le -\beta _{u}^{{}}(x)-q(x)  \hfill\\   
  \quad \quad \quad \le -u_{\max }^{T}R{{u}_{\max }}-q(x)  \hfill\\   
  \quad \quad \quad  \le -{{u}^{T}}Ru-q(x)  \hfill\\  
 \end{gathered} 
\label{proof5}
\end{align}

\noindent
Integrating both sides of (\ref{proof5}) on the time interval $[0,t]$ yields

\begin{align}
\begin{gathered}
  V(x(0))-V(x(t))\ge \int_{0}^{t}{\,[{{u}^{T}}Ru+q(x)}]d\tau
 \end{gathered} 
\label{proof6}
\end{align}

\noindent Let $t\to \infty$, then

\begin{align}
\begin{gathered}
  V(x(0))\ge \underset{t\to \infty }{\mathop{\lim }}\ \int_{0}^{t}{\,[{{u}^{T}}Ru+q(x)}]d\tau= \int_{0}^{\infty}{\,[{{u}^{T}}Ru+q(x)}]d\tau=J(x,u)
 \end{gathered} 
\label{proof7}
\end{align}

\noindent
which implies the optimal controller provides a suboptimal performance (i.e., an upper bound) for the original cost function (\ref{eq4}) and (\ref{eq5}). Next, we show the uniqueness of the solution to (\ref{eq10}). Given $u$, assume there is another Lyapunov function ${{V}^{a}}$ satisfies HJB equation                                                                  
\vspace{-0.05cm}
\begin{equation}
\overline{H}\text{(}{{V}^{a}}\text{)=}0,\text{  }x\in \mathcal{X}.
\label{eq71}
\end{equation}
\vspace{-0.4cm}

\noindent
Since $\overline{r}\left( x,u \right)>0$, we have $\nabla {{V}^{T}}\left( x \right)\left( f\left( x \right)+g\left( x \right)u(x) \right)<0$, $\forall x\in \mathcal{X}\backslash \{0\}$. Subtracting $\overline{H}({{V}^{o}})$ from $\overline{H}({{V}^{a}})$ yields

\vspace{-0.4cm}
\begin{equation}
\overline{H}({{V}^{a}})-\overline{H}({{V}^{o}})={{[\nabla {{V}^{a}}\left( x \right)-\nabla {{V}^{o}}\left( x \right)]}^{T}}\left( f\left( x \right)+g\left( x \right)u(x) \right)=0.
\label{eq72}
\end{equation}
\vspace{-0.2cm}

\noindent
Therefore,  ${{V}^{a}}={{V}^{o}}+\varepsilon$ for some scalar $\varepsilon $,$\forall x\in \mathcal{X}\backslash \{0\}$. Since $f\left( x \right)+g\left( x \right)u(x)\ne 0$, $\forall x\in \mathcal{X}\backslash \{0\}$. Here ${{V}^{a}}(0)={{V}^{\text{*}}}(0)=0$ result in $\varepsilon =0$ and therefore ${{V}^{a}}(x)={{V}^{o}}(x)$ holds for all $\forall x\in \mathcal{X}$. This contradicts with our assumption that there is another ${{V}^{a}}$ satisfy HJB equation.  So, ${{V}^{o}}(x)$ is a unique solution for equation $\overline{H}\left( {{V}^{o}} \right)= 0$.

$\hfill$ $\square$

\smallskip

Note that under Assumption 1, there exists a well-defined Lyapunov function ${{V}^{0}}\in P$ and a control policy ${{u}^{1}} $such that

\vspace{-0.45cm}
\begin{equation}
L({{V}^{0}},{{u}^{1}})=-{{(\nabla {{V}^{0}})}^{T}}\left( x \right)\left( f\left( x \right)+g\left( x \right){{u}^{1}} \right)-\overline{r}\left( x,{{u}^{1}} \right)\ge 0\,\,\,\,
\label{eq26}
\end{equation}                              

\noindent
{\textbf{Remark 2.} Note that in general the optimal control problem does not necessarily have a smooth value function. However, under some mild assumptions \cite{a30}, the value function satisfies premier properties like continuity and continuous differentiability. In this paper, all derivations are performed under the assumption of the existence of a smooth solution to (\ref{eq10}). If the smoothness assumption is relaxed, then one needs to use the theory of viscosity solutions \cite{a30} to find a solution. Note, however, that the existence of the disturbance and so addition of the extra term $\beta(x)$ to the HJB might restrict the class of the systems or the conditions under which the existence of the solution is guaranteed. This is the case for all other approaches that deal with disturbance. For example, for the disturbance-free case, if the system is stabilizable and q(x) is positive definite, then the existence of a unique solution to the HJB equation is guaranteed. However, if $H_{\infty}$ control is employed for disturbance attenuation, then the solution to its corresponding Hamilton-Jacobi-Issac (HJI) equation is not guaranteed under the same conditions and extra conditions on the performance parameters and the disturbance attenuation level are required to guarantee existence of a solution. 

\vspace{-0.4cm}

\subsection{Safety Assurance Using Control Barrier Certificate}
\vspace{-0.1cm}
It is of vital importance for a safety critical system to prevent its state from entering some certain unsafe regions. To design a safe controller, the concept of control barrier function (CBF) can be used. Consider the nonlinear system (\ref{eq1}) with $x\in \mathcal{X}$ , where $\mathcal{X}$ is the allowable set for system’s states, and let  ${{\mathcal{X}}_{\mathbf{0}}}\in \mathcal{X}\in {{\mathbb{R}}^{n}}$ and ${{\mathcal{X}}_{u}}\in \mathcal{X}\in {{\mathbb{R}}^{n}}$ be initial set and unsafe set, respectively. Let there exists a continuously differentiable function $h\in {{C}^{1}}(\mathcal{X}):{{\mathbb{R}}^{n}}\to \mathbb{R}$ such that

\vspace{-0.3cm}
\begin{align}
\begin{gathered}
h\left( x \right)\ge 0,\,\,\forall x\in {{\mathcal{X}}_{0}}, \hfill \\
  h\left( x \right)<0,\,\,\forall x\in {{\mathcal{X}}_{u}}. \hfill \\ 
\end{gathered} 
\label{eq14}
\end{align}
\noindent
\vspace{-0.3cm}

Under disturbance, one must guarantee that regardless of the disturbance value, the system never enters the unsafe set ${{\mathcal{X}}_{u}}$. The input-to-state safety and the robust CBF defined below provide the conditions under which the system trajectories never enter an unsafe set despite the disturbances (as inputs to the system).  \vspace{3pt}

\noindent
\textbf{Definition 1\cite{a24}.} Let ${{\mathcal{X}}_{u}}$ be the unsafe set. The system (\ref{eq1}) is input-to-state safe (ISSf) if there exist $\alpha ,\phi \in \kappa \kappa $ and a strictly increasing function $\sigma$ such that

\vspace{-0.3cm}
\begin{equation}
\sigma (||x(t)|{{|}_{{{\mathcal{X}}_{u}}}})\ge \alpha (||x(t)|{{|}_{{{\mathcal{X}}_{u}}}},t)-\phi ((||u|{{|}_{{{L}^{\infty }}}},t))
\label{eq15}
\end{equation}

\vspace{-0.15cm}

\noindent
holds $\forall t$.

Let the safe set $\mathcal{C}$ be defined as 
\vspace{-0.0cm}
\begin{align}
\begin{gathered}
 \partial \mathcal{C}=\{x\in \mathcal{X}:h(x)=0\},  \hfill \\
  \mathcal{C}=\{x\in \mathcal{X}:h(x)\ge 0\},  \hfill \\ 
  Int(\mathcal{C})=\{x\in \mathcal{X}:h(x)>0\}.  \hfill \\ 
\end{gathered} 
\label{eq16}
\end{align}
\noindent
\vspace{-0.3cm}
\smallskip

\noindent
Then, according to \cite{a24}, in the presence of disturbance $d\ne 0$, $h(x)$ is called a Zeroing CBF (ZCBF) if there exists an extended class $\kappa$ function $\varpi \in \kappa $  that satisfies

\vspace{-0.3cm}
\begin{align}
\begin{gathered}
~\underset{u\in U}{\mathop{\sup }}\,\left\{ \nabla h( x \right)f\left( x \right)+\text{ }\!\!~\!\!\text{ }\nabla h\left( x \right)g\left( x \right)u \hfill \\ 
-\nabla h\left( x \right)g(x){{(\nabla h\left( x \right)g(x))}^{T}} 
+\varpi \left( h\left( x \right) \right)\}  \ge 0,\text{ }\!\!~\!\!\text{  }\!\!~\!\!\text{  }\!\!~\!\!\text{ }\!\!~\!\!\text{ }\quad x\in \mathcal{X} \hfill 
\end{gathered} 
\label{eq17}
\end{align} 

\smallskip

Based on ZCBF $h\left( x \right),$ the safe control space $S\left( x \right)~$ is defined as 

\vspace{-0.3cm}
\begin{align}
\begin{gathered}
S( x )=\{ u\in U|\nabla h( x )f( x )+\text{ }\!\!~\!\!\text{ }\nabla h( x )g( x )u \hfill \\ 
-\nabla h( x)g(x){{(\nabla h( x )g(x))}^{T}}+\varpi ( h( x ) )\ge 0 \}\} ,\text{ }\!\!~\!\!\text{ }x\in \mathcal{X}. \hfill 
\end{gathered} 
\label{eq18}
\end{align} 

The following theorem shows how to design a controller using the concept ZCBF to guarantee that the safe set $\mathcal{C}$ is forward invariant and thus the system is safe.

\smallskip

\noindent
$\textbf{Assumption 3\cite{a24}.}$ The admissible control space $S\left( x \right)$ is path-connected and nonempty. That is, 

\vspace{-0.1cm}
\begin{equation}
Int(S)\ne 0\,\,\,\,and\,\,\,\overline{Int(S)}=S.
\label{eq19}
\end{equation}

\smallskip

\noindent
\textbf{Theorem 2 \cite{a25}.} Consider the set $\mathcal{C}\subset {{\mathbb{R}}^{n}}$ defined in  (\ref{eq16}) and let the ZCBF $h$ be defined as (\ref{eq17}). Then, under Assumption 3,  any Lipschitz continuous controller $u$ such that $u\in S\left( x \right)$  renders the system (\ref{eq1}) ISSf and the safe set $\mathcal{C}$ forward invariant.                $\hfill$ $\square$   \vspace{4pt}

Conditions (\ref{eq16}) and (\ref{eq17}) guarantee that if the system starts from any initial condition $x\in {{\mathcal{X}}_{0}}$ within the safe set, its future trajectories will not enter the unsafe region $x\in {{\mathcal{X}}_{u}}$ for any disturbances. This is because the condition (\ref{eq17}) makes the safe set $\mathcal{C}$ robustly invariant.

\vspace{-0.4cm}

\subsection{Satisficing Control Design}
\vspace{-0.1cm}
Satisficing (good enough) decision-making framework, originated from economical science \cite{a15}, and adopted in control society later \cite{a16,a17}, defines two utility functions: the selectability ${{p}_{x}}(u,x)$ and rejectability ${{p}_{r}}(u,x)$. One then seeks to find a strategy $u$ from a so-called satisficing set defined as

\vspace{-0.2cm}
\begin{equation}
{{S}_{B}}(x)=\left\{ u\in {{\mathbb{R}}^{m}}:{{p}_{s}}(u,x) \succeq B(x){{p}_{r}}(u,x) \right\}
\label{eq20}
\end{equation}

\noindent
where $B(x)$ is called aspiration level. Any control strategy that has a lager selectability index than rejectability index belongs to the satisficing set and is considered to have a good enough performance. It was shown in \cite{a16} that if ${{p}_{s}}(u,x)=-{{(\nabla V)}^{T}}\left( x \right)\left( f\left( x \right)+g\left( x \right)u \right)$ where $V$ is a control Lyapunov function and ${{p}_{r}}(u,x)=r(x,u)$ with $r(x,u)$ defined in (\ref{eq5}), then satisficing controllers (\ref{eq20}) are stabilizing. Moreover, ${{p}_{s}}(u,x)$ indicates the stability index of the system and ${{p}_{r}}(u,x)$ indicates the cost of implementing the controller. It was also shown in \cite{a16} how to parametrize the set of all satisficing controllers. However, it is not clear how to choose one control strategy from the set of satisficing controllers to be applied to the system. We design a policy iteration (PI) algorithm that selects a satisficing strategy that optimizes a performance.
\vspace{-0.3cm}

\section{A SATISFICING SAFE CONTROL DESIGN SCHEME  }

In this section, a novel satisficing safe control framework is designed that guarantees robust safety and a good enough performance. First, a relaxed robust stabilizing control framework is presented in 3.2, inspired by \cite{a26,a27,a28,a29}, in which an infinite-dimensional linear program (LP) is derived to find a robust suboptimal controller. The infinite-horizon LP is then transformed into a sum-of-squares (SOS) program. This framework will then be integrated with CBF in 3.3 to certify robust safety of the resulting controller. 

\vspace{-0.15cm}

\subsection{A Satisficing Control Framework for Safe control with Guaranteed Performance}

We now define a satisficing control framework that defines the selectability index as robust stability and robust safety of controllers and rejectability index as their cost. Consider the system (\ref{eq1}) with the performance function (\ref{eq6}) and (\ref{eq7}). Let $V$ be a control Lyapunov function and $h\left( x \right)$ be a control barrier function. Define

\begin{equation}
\left \{ \begin{array}{l}
{{{p}_{s}}(u,x)} = [{-{{(\nabla V)}^{T}}\left( x \right)\left( f\left( x \right)+g\left( x \right)u \right)}, \quad  \hfill \\ {\nabla h\left( x \right)f\left( x \right)+\text{ }\!\!~\!\!\text{ }\nabla h\left( x \right)g\left( x \right)u-\nabla h\left( x \right)g(x){{(\nabla h\left( x \right)g(x))}^{T}}}]^{T}\\
\\
{{{p}_{r}}(u,x)} = [{\overline{r}\left( x,u \right)}, \quad {-\eta \varpi \left( h\left( x \right) \right)}]^{T}
\end{array}\right.
\label{eq21}
\end{equation}

\noindent 
where $\overline{r}\left( x,u \right)$ is defined in (\ref{eq7}), $\eta >0$ is a constant, ${{p}_{s}}(u,x)$ is selectability and is associated with the robust stability and robust safety of the system, and ${{p}_{r}}(u,x)$ is rejectability and is associated with the controller cost and safety aggressiveness.  Define now the satisficing set as (\ref{eq20}) where $B(x)=diag({{b}_{1}}(x), \,\,\,{{b}_{2}}(x))$ with ${{b}_{1}}(x)$ as the aspiration level on stability and ${{b}_{2}}(x)$ the aspiration level on safety. \vspace{2pt}

\noindent
\textbf{Remark 3.} The parameter $\eta $ in the rejectibility index of (\ref{eq21}) reveals how rapidly $\varpi \left( h\left( x \right) \right)$ damps as the system’s states get further away from the safe boundaries and thus it is treated as a rejectability index. A larger $\eta $ increases system’s aggressiveness in favor of having more flexibility in the safe set. The aspiration level reflects the expectation of the designer on satisfaction of the controller, and is also related to the solution space in the satisficing set ${{S}_{B}}(x)$. 

The next theorem shows that under Assumptions 1-3, there is always a feasible solution to (\ref{eq20}), (\ref{eq21}) by choosing an appropriate aspiration level. The feasible solution makes the system both robustly safe and robustly stable if there is no conflict between safety and stability and sacrifices on the stability in case of a conflict.

\smallskip

\noindent
{\textbf{Theorem 3.} Let Assumptions 1 and 3 hold. Then, there exists an aspiration level $B(x)$ for which the satisficing set (\ref{eq20}) with ${{p}_{s}}(u,x)$ and ${{p}_{r}}(u,x)$ defined in (\ref{eq21}) has a feasible solution. \vspace{2pt}

\noindent
\textit{Proof.} By Assumption 3, there is always a safe control policy that satisfies the second inequality in (\ref{eq20}), (\ref{eq21}). On the other hand, by Assumption 1, there is a stabilizing controller that satisfies the first constraints. If there is no conflict between safety and stability, then there is a controller that satisfies both inequalities in (\ref{eq20}), (\ref{eq21}), the satisficing set is nonempty. Assume now that there is a conflict between two constraints in (\ref{eq20}), (\ref{eq21}). Let

\vspace{-0.25cm}
\begin{equation}
{{b}_{1}}(x)=\frac{\overline{r}\left( x,u \right)+\text{ }\!\!\delta\!\!\text{ }}{\overline{r}\left( x,u \right)},\,\,\,\,{{b}_{2}}(x)=\theta 
\label{eq23}
\end{equation}

\noindent 
where $\text{ }\!\!\delta\!\!\text{ }$ is a function of $x$ representing the conflict between two constraints, and $\theta >0$ is the aspiration on the safety level. Based on (\ref{eq17}), the safety constraint is satisfied by choosing any $\theta >0$ . On the other hand, based on the defined ${{b}_{1}}(x)$ in (\ref{eq23}), the first constraints in the equality ${{p}_{s}}(u,x)=B(x){{p}_{r}}(u,x)$ becomes

\vspace{-0.4cm}
\begin{equation}
-{{(\nabla V)}^{T}}\left( x \right)\left( f\left( x \right)+g\left( x \right)u \right)-\overline{r}\left( x,u \right)=\text{ }\!\!\delta\!\!\text{ }.
\label{eq24}
\end{equation}
\vspace{-0.35cm}

\noindent
Therefore, any aspiration level of stability ${{b}_{1}}^{'}(x)$ satisfying $\text{ }{{b}_{1}}^{'}(x)\leq \text{ }{{b}_{1}}(x)$ also satisfies safe constraint as

\vspace{-0.3cm}
\begin{equation}
-{{(\nabla V)}^{T}}\left( x \right)\left( f\left( x \right)+g\left( x \right)u \right)\ge -\theta \eta \varpi \left( h\left( x \right) \right).
\label{eq25}
\end{equation}
\vspace{-0.45cm}

\noindent 
This completes the proof.  $\hfill$ $\square$                                                                                                                             

Equation (\ref{eq24}) shows that $\delta$ indicates the sacrifice on stability and performance, since (\ref{eq25}) resembles the Bellman inequality and will be used later to improve performance. This function will be optimized later to minimize the sacrifice on performance as much as possible. Note that a stabilizing solution to (\ref{eq25}) is guaranteed under Assumption 1 when $\delta=0$. When $\delta$ is nonzero, however, the solution to (\ref{eq25}) might not be stabilizing and this can occur only when the safety is in conflict with stability. In this paper, instead of parameterizing the set of all satisficing controllers, a PI algorithm is designed to optimize over all satisficing control policies by relating the control Lyapunov function   with the Bellman inequality solution, which is a performance-oriented control Lyapunov function and can be iteratively optimized while assuring that every improved policy remains in the satisficing set and thus guarantees stability. To this end, it is shown in the next subsection how to optimize over the set of satisficing control policies by iteratively solving Bellman inequalities while ignoring the safety constraint. 3.3 combines safety constraints and bellman inequalities to guarantee safety of improved policies.

\vspace{-0.4cm}

\subsection{Relaxed Robust Stabilizing Optimal Control Design}
\vspace{-0.1cm}
In this subsection, a relaxed robust optimal control framework is presented. While safety is ignored here, this framework allows incorporating safety constraints, as shown in the next subsection. Inspired by \cite{a26,a27,a28,a29}, a finite-dimensional linear program (LP) is first derived. This finite-dimensional optimization problem is integrated with CBF to include safety and is solved using SOS in the next subsection.

\smallskip

\noindent
{\textbf{Problem 1 (Relaxed suboptimal robust stabilizing control design problem) } 

Consider the system (\ref{eq1}) with the performance index (\ref{eq6}), (\ref{eq7}). Find the value function $V$ by solving

\vspace{-0.25cm}
\begin{align}
\begin{gathered}
  \underset{V}{\mathop{\min }}\,\mathop{\int }_{\Omega }^{{}}V\left( x \right)dx \\
  \text{s.t}.\quad \text{  }\!\!~\!\!\text{  }\!\!~\!\!\text{   }\!\!~\!\!  \overline{H}\left( V \right)\le 0  \\
   V\in P  \\ 
\end{gathered} 
\label{eq27}
\end{align}

\noindent
where $\Omega $ is the area in which system performance is expected to be improved, and $\overline{H}(V)$ is defined in (\ref{eq11}).

This is inspired by \cite{a26,a27,a28} in which it is shown that an optimal control problem can be transformed into an infinite-dimensional LP. Instead of searching for optimal value function over whole an infinite-dimensional space, in \cite{a28}, the optimization is performed over an interested region $\Omega $, which makes the problem finite-dimensional and tractable. In contrast to \cite{a28},  a modified HJB inequality $\overline{H}\left( V \right)<0$ with the modified reward function is used to guarantee robust stability. This framework will allow us to incorporate safety constraints later and find satisficing control solutions that satisfy safety and optimize over stabilizing solutions. The following theorem shows some premier properties of Problem 1.

\smallskip

\noindent
{\textbf{Theorem 4.} Consider the system (\ref{eq1}) and let Assumptions 1-3 hold. Then, 

1) Problem 1 has a feasible value function solution. 

2) If $V$ is the unique solution of Problem 1, then, the control solution                      

\vspace{-0.25cm}
\begin{equation}
\overline{u}(x)=-\frac{1}{2}{{R}^{-1}}{{(\nabla {{V}^{T}}\left( x \right)g(x))}^{T}}
\label{eq28}
\end{equation}

\noindent
is globally robust stable and belongs to the satisficing set (\ref{eq20}) when safety is ignored.

3) The control policy (\ref{eq28}) provides an upper bound for the cost (\ref{eq4}).

4) The following inequalities hold for $x\in \mathcal{X}$ along the trajectories of the closed system (\ref{eq1}) with the controller (\ref{eq28}), 

\vspace{-0.25cm}
\begin{equation}
V({{x}_{0}})+\int\limits_{0}^{\infty }{H(V(x(t)))dt}\le {{V}^{o}}({{x}_{0}})\le V({{x}_{0}})
\label{eq29}
\end{equation}
\vspace{-0.2cm}

5) The value function ${{V}^{o}}$ in (\ref{eq10}) is a global robust optimal solution to (\ref{eq27}).

\smallskip
\noindent
\textit{Proof.} See APPENDIX.           $\hfill$ $\square$

  Iterative policy iteration algorithms can be designed to solve Problem 1. The need for safety assurance, however, makes existing policy iteration algorithms invalid, as they cannot guarantee safety. In the next subsection, to find a robust safe control policy with guaranteed performance, safety constraints satisfaction is incorporated by including a CBF as another inequality to Problem 1 and a novel PI algorithm is developed to solve it. Its connection to satisficing controllers is also shown. 

\vspace{2pt}
\noindent
{\textbf{Remark 4.} A feasible solution $V$ to (\ref{eq27}) may not be the true cost function associated with $\overline{u}$. However, it is an upper bound of the practical cost.

\vspace{-0.4cm}

\subsection{Robust Safe Satisficing Control with Performance Guarantee: A Novel Framework}
\vspace{-0.1cm}
While the controller designed by solving Problem 1 guarantees performance and robustness and belongs to satisficing set that only concerns stability, it cannot assure safety. On the other hand, the control design based on the CBF satisfying (\ref{eq20}) guarantees safety but might result in poor performance. To bring the best of both worlds together, in this section, we aim to design robust stabilizing safe controllers that provide guaranteed performance within the volume of the certified safe area.

\smallskip

\noindent
{\textbf{Problem 2 (Satisficing safe control design with performance guarantee) } 

Consider the system (\ref{eq1}) with the performance function (\ref{eq6}), (\ref{eq7}). Find the value function  that solves

\vspace{-0.25cm}
\begin{align}
\begin{gathered}
  \underset{V,\text{ }\!\!\delta\!\!\text{ }}{\mathop{\min }}\,\mathop{\int }_{\mathcal{L}}^{{}}V\text{ }\!\!~\!\!\text{ dx }\!\!~\!\!\text{ }+{{k}_{\text{ }\!\!\delta\!\!\text{ }}}{{\text{ }\!\!\delta\!\!\text{ }}^{2}}  \\
   \text{s.t}.\quad \text{  }\!\!~\!\!\text{  }\!\!~\!\!\text{   }\!\!~\!\!\text{ }\overline{H}\left( V \right)\le \text{ }\!\!\delta\!\!\text{ }   \\
    \text{               }\nabla h\left( x \right)f\left( x \right)+\text{ }\!\!~\!\!\text{ }\nabla h\left( x \right)g\left( x \right)u-\nabla h\left( x \right)g(x){{(\nabla h\left( x \right)g(x))}^{T}}\\ + \varpi \left( h\left( x \right) \right)\ge 0\text{ }\!\!~\!\!\text{ }   \\ 
\end{gathered} 
\label{eq31}
\end{align} 

\vspace{-0.35cm}
\noindent
where $\overline{H}(V)$ is defined in (\ref{eq11}), $\mathcal{L}$ is the interested safe region in which the performance is expected to be improved, ${{\text{k}}_{\text{ }\!\!\delta\!\!\text{ }}}>0$ is a design parameter that trades off between the system’s aggressiveness toward performance and the safety, $\delta $ is a relaxation factor. 

Note that the relaxation factor $\delta $ can be interpreted as the system's aspiration level for the performance that shows how much we sacrifice the performance when both safety and performance cannot be satisfied together. This relaxation factor is minimized to get as much performance as possible.

\smallskip

\noindent
{\textbf{Lemma 1.} The solution to Problem 2 belongs to the satisficing set (\ref{eq20}).

\smallskip
\noindent
\textit{Proof.} Recalling (\ref{eq23}), (\ref{eq24}) and (\ref{eq25}), one can see that the constraints in (\ref{eq31}) can be transformed to (\ref{eq20}) by selecting suitable aspiration levels given in Theorem 4. Therefore, the search for the optimal solution is over the space of satisficing set (\ref{eq20}), and, thus, the solution to Problem 2 belongs to (\ref{eq20}).                                                                                             $\hfill$ $\square$

\smallskip

\noindent
{\textbf{Theorem 5.} The safe optimization problem (\ref{eq31}) has a feasible solution. \vspace{3pt}

\noindent
\textit{Proof.} Based on Theorem 4, a robust safe control policy $u$ exists by selecting a suitable aspiration level. Let write this control policy as $u={{u}^{*}}+{{u}^{safe}}$ where ${{u}^{*}}=-\frac{1}{2}{{R}^{-1}}{{({{(\nabla {{V}^{*}})}^{T}}(x)g(x))}^{T}}$ is part of the control that is used to optimize the performance without concerning safety, and is given by \cite{a31}, and ${{u}^{safe}}$ is added to ${{u}^{*}}$ to guarantee safety. Reformulate now the HJB equation as follows

\begin{align}
\begin{gathered}
 \overline{H}({{V}^{*}})=-\frac{1}{4}{{(\nabla {{V}^{*}})}^{T}}(x)g(x){{R}^{-1}}{{({{(\nabla {{V}^{*}})}^{T}}(x)g(x))}^{T}} \hfill \\ 
  \quad \quad \quad +q(x)+{{(\nabla {{V}^{*}})}^{T}}(x)f(x)+d_{max}^TRd_{max}+\beta_u(x) \hfill \\ 
  \quad \quad \quad ={{(\nabla {{V}^{*}})}^{T}}(x)(f(x)+g(x)u)+\overline{r}(x,u) \hfill \\ 
  \quad \quad \quad ={{(\nabla {{V}^{*}})}^{T}}(x)(f(x)+g(x)u)+\overline{r}(x,u) \hfill \\ 
  \quad \quad \quad +{{({{u}^{*}})}^{T}}R{{u}^{*}}-{{u}^{T}}Ru-{{(\nabla {{V}^{*}})}^{T}}(x)g(x){{u}^{safe}} \hfill \\  
  \quad \quad \quad ={{(\nabla {{V}^{*}})}^{T}}(x)(f(x)+g(x)u)+\overline{r}(x,u) \hfill \\  
  \quad \quad \quad +{{({{u}^{*}})}^{T}}R{{u}^{*}}-{{u}^{T}}Ru+2{{({{u}^{*}})}^{T}}R{{u}^{safe}} \hfill \\ 
  \quad \quad \quad ={{(\nabla {{V}^{*}})}^{T}}(x)(f(x)+g(x)u)+\overline{r}(x,u) \hfill \\  
  \quad \quad \quad +{{({{u}^{*}})}^{T}}R{{u}^{*}}-{{u}^{T}}Ru+2{{({{u}^{*}})}^{T}}R(u-{{u}^{*}}) \hfill \\ 
  \quad \quad \quad ={{(\nabla {{V}^{*}})}^{T}}(x)(f(x)+g(x)u)+\overline{r}(x,u) \hfill \\  
  \quad \quad \quad -{{({{u}^{*}})}^{T}}R{{u}^{*}}-{{u}^{T}}Ru+2{{({{u}^{*}})}^{T}}Ru \hfill \\  
  \quad \quad \quad ={{(\nabla {{V}^{*}})}^{T}}(x)(f(x)+g(x)u)+\overline{r}(x,u) \hfill \\  
  \quad \quad \quad -{{(u-{{u}^{*}})}^{T}}R(u-{{u}^{*}}) \hfill \\ 
  \quad \quad \quad ={{(\nabla {{V}^{*}})}^{T}}(x)(f(x)+g(x)u)+\overline{r}(x,u)-{{({{u}^{safe}})}^{T}}R{{u}^{safe}} \hfill \\  
\end{gathered} 
\label{eq32}
\end{align} 

\noindent
where $\overline{r}$ is defined in (\ref{eq7}) and (\ref{eq9}). While ${{u}^{*}}$ is robust stabilizing, if robust safe control ${{u}^{safe}}$ is in conflict with the robust stability, the overall control input $u$ might not be robust stabilizing, i.e., ${{(\nabla {{V}^{*}})}^{T}}\left( f(x)+g(x)u \right)+\overline{r}\left( x,u \right)>0$ might not be satisfied at some points. By choosing an appropriate slack variable $\delta $ to resolve the conflict between safety and stability, however, one has

\vspace{-0.35cm}
\begin{equation}
\overline{H}({{V}^{*}})\text{-}\delta \text{=}{{(\nabla {{V}^{*}})}^{T}}(x)\left( f(x)+g(x)u \right)+\overline{r}\left( x,u \right)-\delta \le 0
\label{eq33}
\end{equation}

\noindent 
for some $\delta$. On the other hand, since $u$ is safe, based on the converse CBF theorem \cite{a32}, there exists a barrier certificate $h(x)$ satisfying

\vspace{-0.35cm}
\begin{align}
\begin{gathered}
\text{ }\nabla h\left( x \right)f\left( x \right)+\text{ }\!\!~\!\!\text{ }\nabla h\left( x \right)g\left( x \right)u-\nabla h\left( x \right)g(x){{(\nabla h\left( x \right)g(x))}^{T}}  +\varpi \left( h\left( x \right) \right)\ge 0. \hfill\\
\end{gathered} 
\label{eq34}
\end{align}

$\hfill$ $\square$

\noindent
\textbf{Assumption 4\cite{a28}.} There exists a smooth mapping ${{V}^{0}}:{{\mathbb{R}}^{n}}\to \mathbb{R}$, such that ${{V}^{0}}\in \mathbb{R}{{[x]}_{2,2r}}\cap P$ and $L({{V}^{0}},{{u}^{1}})+\delta$ is SOS, where $L({{V}^{0}},{{u}^{1}})=-\overline{H}\text{(}{{V}^{0}}\text{)}$.

Solving optimization Problem 2 is non-trivial in general. If both Bellman inequality and CBF inequality constraints are restricted to SOS constraints, SOS program can be used to significantly reduce the computational burden in finding a solution to this optimization problem. However, since $\overline{H}(V)$ is bilinear in $V$, it makes the optimization problem hard or even impossible to solve using SOS. Therefore, we propose a robust safe policy iteration algorithm that iterates on a Bellman inequality, which is linear in $V$, instead of directly solving for $\overline{H}(V)\le \delta $. Using this Bellman inequality, a policy evaluation step that will find the value function ${V^{i}}$ and corresponding to a robust safe control policy ${{u}^{i}}$, and policy improvement step will find an improved policy ${{u}^{i+1}}$ for which its safety is certified by adding the CBF inequality. We assume that an initial robust safe control policy ${{u}^{0}}$ is given, which can be found by a control policy that only satisfies the CBF without any concern about optimality.

To evaluate a given policy ${{u}^{i}}$, i.e., to find the value function ${{V}^{i}}$ corresponding to it, a Bellman inequality based on the modified reward function must be solved, which requires knowing $\beta_{u^i}(x)$ in the modified reward functions, which in turns require knowing a bound on ${u^i}_{\max }^{T}R \,\, {u^i}_{\max }$, as defined in Theorem 1. Therefore, before the policy evaluation step, we first find this bound and thus $\beta_{u^i}(x)$. Since $u^{i}$, and, consequently $u^{i^T}R \,\, {u^i}$, is polynomial, one can write it as $u^{i^T}R \,\, {u^i}=c_im_i^{x}$ where $m_i^{x}$ is the set of monomials and $c_i$ is the vector of coefficients. To find ${u^i}_{\max }^{T}R \,\, {u^i}_{\max }=max u^{i^T}R \,\, {u^i}$, one can then solve the following optimization problem.

\begin{align}
\begin{gathered}
   {u^i}_{\max }^{T}R \,\, {u^i}_{\max } =max \, \, c_im_i^{x}  \\ 
  \,\,s.t.\,\,\,\,x\,\,\in\,\,\mathcal{X} \\ 
 \end{gathered}
 \label{u_modares1}
\end{align}

\noindent  
However, polynomial optimization is NP-hard and, instead, we obtain a lower bound for ${u^i}_{\max }^{T}R \,\, {u^i}_{\max }$ by solving 

\begin{align}
\begin{gathered}
   L_{p} =min \, \, \gamma  \\ 
  \,\,s.t.\,\,\,\,\gamma\,-\,c_im_i^{x} \, \ge \,0\\ 
 \end{gathered}
 \label{u_modares2}
\end{align}

\noindent 
which is an SOS optimization and can be efficiently solved. Note that ${u^i}_{\max }^{T}R \,\, {u^i}_{\max }\leq L_{p}$ and if we choose $\beta_{u^i}(x)=L_{p}$, then $\beta_{u^i}(x)\ge (u^i_{max})^T \,\ R \,\ u^i_{max}$ which satisfies the condition of reward function in Theorem 1. Based on this optimization, the following policy evaluation step is proposed.

\vspace{2pt}

\noindent
\textbf{Safe policy evaluation step:} Given a robust safe control policy ${{u}^{i}}$, find the bound for $u^{i}_{max}$ using \eqref{u_modares2} and then find ${{V}^{i}}$ and ${{\delta }_{i}}$ that solve the following optimization problem:

\vspace{-0.35cm}
\begin{align}
\begin{gathered}
  \underset{{{\text{V}}^{i}},{{\delta }^{i}}}{\mathop{\min }}\,\mathop{\int }_{\mathcal{L}}^{{}}{{V}^{i}}\text{ }\!\!~\!\!\text{ dx }\!\!~\!\!\text{ +}{{\text{k}}_{\delta }}{{\delta }_{i}}^{2} \\
  \text{s.t}. \quad \text{ }\!\!~\!\! L({{V}^{i}},{{u}^{i}})=-{{(\nabla {{V}^{i}})}^{T}}\left( x \right)\left( f\left( x \right)+g\left( x \right){{u}^{i}} \right)-\overline{r}\left( x,{{u}^{i}} \right)\ge -{{\delta }_{i}}  \\
   {{V}^{i-1}}-{{V}^{i}}\ge 0   \\ 
\end{gathered} 
\label{eq35}
\end{align} 

In terms of SOS, this optimization problem is transformed into

\vspace{-0.35cm}
\begin{align}
\begin{gathered}
  \underset{{{\text{V}}^{i}},{{\delta}^{i}}}{\mathop{\min }}\,\mathop{\int }_{\mathcal{L}}^{{}}{{\text{V}}^{i}}\text{ }\!\!~\!\!\text{ dx }\!\!~\!\!\text{ +}{{\text{k}}_{\delta }}{{\delta }_{i}}^{2}  \\
  \text{s.t}.\quad \text{ }\!\!~\!\! L({{V}^{i}},{{u}^{i}})+{{\delta }_{i}}\text{ }\!\!~\!\!\,\,\text{is}\,\,\text{ }\!\!~\!\!\text{SOS}, \text{ }\!\!~\!\! \quad \forall x\in \mathcal{X}  \\
   {{V}^{i-1}}-{{V}^{i}}\,\,\,\text{is}\,\,\,\text{SOS}  \\ 
\end{gathered} 
\label{eq36}
\end{align} 

\noindent
where $V={{p}^{T}}{{\overrightarrow{m}}_{2,2r}}(x)$, and ${{V}_{i}}=p_{^{i}}^{T}{{\overrightarrow{m}}_{2,2r}}(x)$ . 

In the policy evaluation step (\ref{eq35}) , the value function corresponding to a given policy is found while minimizing the relaxation factor ${{\delta }_{i}}$. Note that since a robust safe control policy ${{u}^{i}}$ might not necessarily be robust stabilizing, therefore, $L({{V}^{i}},{{u}^{i}})$ might not be positive semidefinite. \vspace{3pt}

\noindent {\textbf{Remark 5.} Instead of performing two SOS optimization to evaluate a given policy, (one  for finding $u^i_{max}$ in each step), one can regard every element of $u^i_{max}=[u^i_1,...,u^i_m]$ as a decision variable and incorporate it through the following optimization problem.

\begin{align}
\begin{gathered}
   \min \,{u^i_{j_{\max }}} \\ 
  \,\,s.t.\,\,\,\,{{u^i_j}_{\max }}-{{u}^i_j}\,\,is\,\,SOS \\ 
 \end{gathered}
 \label{u_max1}
\end{align}

\noindent
where the $u^{i^j}$ is the $j-th$ element of the improved safe policy $u^i$ found in Step 3. Then, ${u^i}_{\max }^{T}R \,\, {u^i}_{\max }$ can be calculated since ${u^i}_{\max }=[u^i_{1_{\max} },...,u^i_{m_{\max }}]$. Alternatively, the $u^{i}_{j_{max}}$ can be defined as a polynomial which is desired to be minimized in a domain of interests. That is,

\begin{align}
\begin{gathered}
   \underset{{u}^{i}_{j_{\max} }}{\mathop{\min \,}}\,\int_{\mathcal{D}}{{u^{i}_{j_{\max }}}} \\ 
  \,\,s.t.\,\,\,\,{{u}^{i}_{j_{\max }}}-{u}^{i}\,\,is\,\,SOS \\ 
   \end{gathered}
 \label{u_max2}
 \end{align}
 
 \vspace{0.35cm}

 \noindent 
 where $\mathcal{D}$ is the domain of the interested in which the $u^{i}_{max}$ is desired to be minimized. Incorporating these extra SOS constraints into the policy improvement step \eqref{eq35} relaxes the requirement of solving the SOS optimization \eqref{u_modares2}. 
 
 Once a policy is evaluated, an improved control policy with ISSf certification is found. The following lemma shows how to find a safety-certified improved control policy.

\smallskip

\noindent
{\textbf{Lemma 2 (Safe policy improvement).}  Let ${{u}^{i}}$ be a robust safe control policy with guaranteed value function ${{V}^{i}}$. Then, an improved safety certified control policy ${{u}^{i+1}}$ can be found by solving following optimization problem

\vspace{-0.35cm}
\begin{align}
\begin{gathered}
  \underset{{{u}^{safe}},Z}{\mathop{\min }}\,\,{{({{u}^{safe}})}^{T}}R{{u}^{safe}}   \\
 \text{s.t}. \quad \text{ }\!\!~\!\! {{u}^{i+1}}={{u}^{safe}}-\frac{1}{2}{{R}^{-1}}{{(\nabla {{V}^{i}}(x)g(x))}^{T}}   \\
   \nabla {{h}}\left( \text{x} \right)f\left( x \right)+\text{ }\!\!~\!\!\text{ }\nabla {{h}}\left( \text{x} \right)\text{g}\left( \text{x} \right){{u}^{i+1}}+Z\,{{h}}\left( x \right)\\ -\nabla {{h}}\left( x \right)g(x){{(\nabla {{h}}\left( x \right)g(x))}^{T}}\,\,\text{is }\,\,\text{SOS }\!\!~\!\!\text{ }  \\ 
   Z \text{ }\!\!~\!\!\,\,\text{is }\,\,\text{SOS}  \\
\end{gathered} 
\label{eq37}
\end{align}

\noindent
\textit{Proof.} To find an improved control policy, we use stationary condition \cite{a33} that minimizes the Bellman equation while satisfying the CBF and find an improved policy. Note that the Bellman equation can be written as

\begin{align}
\begin{gathered}
 L({{V}^{i}},{{u}^{i+1}})=-{{(\nabla {{V}^{i}})}^{T}}\left( x \right)\left( f\left( x \right)+g\left( x \right){{u}^{i+1}} \right)-\overline{r}\left( x,{{u}^{i+1}} \right) \hfill\\ 
 \qquad \qquad \text{ }\!\!~\!\!= -{{(\nabla {{V}^{i}})}^{T}}\left( x \right)\left( f\left( x \right)+g\left( x \right){{({{u}^{opt}})}^{i+1}} \right)\hfill\\ 
  \quad \qquad \qquad \text{ }\!\!~\!\!+{{({{u}^{safe}})}^{T}}R{{u}^{safe}}-\overline{r}\left( x,{{({{u}^{opt}})}^{i+1}} \right) \hfill\\
\end{gathered} 
\label{eq38}
\end{align} 

\noindent
where ${{u}^{i+1}}={{u}^{opt}}+{{u}^{safe}}$. Minimizing the term $-{{(\nabla {{V}^{i}})}^{T}}\left( x \right)\left( f\left( x \right)+g\left( x \right){{({{u}^{opt}})}^{i+1}} \right)-\overline{r}\left( x,{{({{u}^{opt}})}^{i+1}} \right)$ using stationarity condition results in ${{u}^{opt}}=-\frac{1}{2}{{R}^{-1}}{{({{(\nabla {{V}^{i}})}^{T}}(x)g(x))}^{T}}$. Therefore, minimizing ${{({{u}^{safe}})}^{T}}R{{u}^{safe}}$ as the second term while setting ${{u}^{i+1}}={{u}^{safe}}-\frac{1}{2}{{R}^{-1}}{{({{(\nabla {{V}^{i}})}^{T}}(x)g(x))}^{T}}$ optimizes the performance. Since the control must certify safety constraint, the CBF inequality must also be considered.                  $\hfill$ $\square$

\vspace{2pt}
\noindent
{\textbf{Remark 6.} In (\ref{eq36}) and (\ref{eq37}), $\delta$ and ${{u}^{safe}}$ are polynomials which can be written in the form of Square Matrix Representation (SMR) as ${{P}^{T}}(x)QP(x)$, where $P(x)$ is a vector of monomials, and $Q$ is a symmetrical coefficient matrix. In order to solve this optimization problem, we adopt a typical way in the literature \cite{a34} to minimize the $trace(Q)$ to get smaller $\delta$ and ${{u}^{safe}}$ for objective function in (\ref{eq36}) and (\ref{eq37}). The SOS program (\ref{eq37}) involves bilinear decision variables. It can be solved efficiently by splitting into several smaller SOS programs as presented in \cite{safe1}.


\smallskip
\noindent
{\textbf{Remark 7 \cite{barr}.} To find a feasible $h(x)$ for performing the policy improvement step, one can solve the following optimization problem.

\begin{align}
\begin{gathered}
   \,\,\,\,\,\,\,\,Find\,\,\,h(x),\,\,{{\sigma }_{1}}(x)\,and\,\,{{\sigma }_{2}}(x)\, \\ 
  s.t.\,\,\,h(x)-\varepsilon {{\sigma }_{1}}(x){{x}_{0}}(x)\,\,is\,\,SOS \\ 
  \,\,\,\,\,\,-\,h(x)-\varepsilon {{\sigma }_{1}}(x){{x}_{o}}(x)\,\,is\,\,SOS \\ 
  \,\,\,\,\,\,\,\,\,\,\,\,\,\,\,{{\sigma }_{1}}(x)\,and\,\,{{\sigma }_{2}}(x)\,\,is\,\,SOS \\ 
   \end{gathered}
  \label{findh}
\end{align}

\noindent
However, there might be multiple $h(x)$ solutions. By maximizing the margin $\varepsilon$ of the barrier certificate constraint, an optimal option of $h(x)$ can be obtained. This method enlarges the feasible solution space of ${{u}^{safe}}$ for optimizing in successive step 1 which speeds up the convergence of optimization procedure \cite{a34,safe1}.

\smallskip

We now combine the safe policy evaluation step (\ref{eq36}) and safe policy improvement step performed in \eqref{eq37}, the following algorithm is presented to iteratively solve Problem 2.

\smallskip
\smallskip

 \noindent\hrulefill\\
 {\bf Algorithm 1: Satisficing safe control design framework.}  \\
 \vspace{.02in}
 {\noindent\hrulefill
 \vspace{.01in}
  \normalsize
 	\begin{algorithmic}[1]
 		\State Initialize with $({V^0},{u^1})$ that satisfies Assumption 4 and set a sum of square variable $\neg $ as a threshold variable.
 	\Procedure{}{} $\forall \,i=1,2,...,N$
 		\State Given $u^{i}$, let ${{V}_{i}}=p_{^{i}}^{T}{{\overrightarrow{m}}_{2,2r}}(x)$, calculate the value function $V^{i}$ and the relaxation variable variable $\delta_{i}$ using (\ref{eq36}) through SOS program. Then, calculate and justify whether $-({V^{i - 1}} - {V^i}) + \neg \,\,\,{\rm{is}}\,\,\,{\rm{SOS}}$. The algorithm stops if $-({V^{i - 1}} - {V^i}) + \neg \,\,\,{\rm{is}}\,\,\,{\rm{SOS}}$, otherwise go to Step 4.
 		\State Iteratively search for an improved policy $u^{i+1}$ using \eqref{eq37}. Then, use $u^{i+1}$ into Step 3 to calculate a new value function.
 	\EndProcedure
 	\vspace{.01in}	
 	\end{algorithmic}
 	\hrulefill
 \normalsize\\

It should be noticed that in Step 3, the condition $-({V^{i - 1}} - {V^i}) + \neg \,\,\,{\rm{is}}\,\,\,{\rm{SOS}}$  implies $|{V^{i - 1}} - {V^i}|\, < \neg \,$. More specifically, Algorithm 1 terminates when the value function $V^{i}$ stops decreasing.

\smallskip
\noindent
{\textbf{Remark 8.} The presented robust safe satisficing control scheme integrates barrier certificate with performance-driven Lyapunov to assure safety while sacrificing as little as possible on performance.

\smallskip

\noindent
{\textbf{Theorem 6.} Consider Assumptions 1-4 for the system (\ref{eq1}).  Then,

1) The policy evaluation step (\ref{eq35})  has a nonempty feasible set. 

2) The closed-loop system (\ref{eq1}) with the controller  ${{u}^{i}}$ derived after each safe policy iteration is robust safe and guarantee the globally robust stability as much as possible.

3) There exists a positive definite ${{V}^{*}}\in \mathbb{R}{{[x]}_{2,2r}}$ such that for any ${{x}_{0}}\in D$, ${{V}^{*}}({{x}_{0}})\le {{V}^{i}}({{x}_{0}})$ holds. Besides, $\underset{i\to \infty }{\mathop{\lim }}\,{{V}^{i}}({{x}_{0}})\to {{V}^{*}}({{x}_{0}})$. 

4) Along the solution of system with $u^{*}(x)=-\frac{1}{2}{{R}^{-1}}{{({{(\nabla {{V}^{*}})}^{T}}\left( x \right)g(x))}^{T}}$, the following holds

\vspace{-0.25cm}
\begin{equation}
V^{*}({{x}_{0}})+\int\limits_{0}^{\infty }{H(V^{*}(x(t)))dt}\le {{V}^{o}}({{x}_{0}})
\label{eq50}
\end{equation}
\vspace{-0.2cm}

\noindent
\textit{Proof:}

1) The following mathematical induction steps are used to prove part 1.

i) Suppose $i=1$. Then, under Assumption 4, $L({{V}^{0}},{{u}^{_{1}}})+\delta $ is SOS. Therefore, $V={{V}^{0}}$ is a feasible solution to (\ref{eq35}).

ii) Assume now that $V={{V}^{j-1}}$ is an optimal solution to the (\ref{eq35}) with $i=j-1>1$. In the following, it is show that $V={{V}^{j-1}}$ is then a feasible solution to the same problem with $i=j$.          

From the safe policy improvement step (\ref{eq37}), by definition,  ${{u}^{i}}={{u}^{safe}}-\frac{1}{2}{{R}^{-1}}{{(\nabla {{V}^{i-1}}(x)g(x))}^{T}}$ and

\begin{align}
\begin{gathered}
L({{V}^{j-1}},{{u}^{_{j}}})+\delta =-{{(\nabla {{V}^{j-1}})}^{T}}(f(x)+g(x){{u}^{_{j}}}+g(x)\omega)-\bar{r}(x,{{u}^{_{j}}})+\delta   \hfill\\   =L({{V}^{j-1}},{{u}^{_{j-1}}})+\delta -{{(\nabla {{V}^{j-1}})}^{T}}g(x)({{u}^{_{j}}}-{{u}^{_{j-1}}})+{{(u_{^{^{^{{}}}}}^{^{j-1}})}^{T}}Ru_{^{^{^{{}}}}}^{^{j-1}}+\beta_{u^{j-1}}(x)-\beta_{u^{j}}(x) -{{(u{^{j}})}^{T}}Ru_{^{^{^{{}}}}}^{^{j}}    \hfill\\
  = L({{V}^{j-1}},{{u}^{_{j-1}}})+{{({{u}^{_{j-1}}}-{{u}^{j}})}^{T}}R({{u}^{_{j-1}}}-{{u}^{j}})+\delta +\beta_{u^{j-1}}(x)-\beta_{u^{j}}(x -2{{({{u}^{safe}})}^{T}}R({{u}^{_{j}}}-{{u}^{_{j-1}}})  \hfill \\
\end{gathered} 
\label{eq42}
\end{align} 

Under the induction assumption, one has ${{V}^{_{j-1}}}\in \mathbb{R}{{[x]}_{2,2r}}$ and $L({{V}^{j-1}},{{u}^{_{j-1}}})+\delta$ is SOS. By selecting a suitable relax variable $\delta$, the affect of safe policy ${{u}^{safe}}$ on positivity of (\ref{eq42}) is eliminated. Hence, $L({{V}^{j-1}},{{u}^{_{j}}})+\delta$ is SOS. As a result, ${{V}^{j-1}}$ is a feasible solution to the SOS program (\ref{eq35}) with $i=j$.

2) We fist show that ${{u}^{i}}$ derived after each safe policy improvement is robustly safe. It can be seen from Algorithm 1 that  $u={{u}^{opt}}+{{u}^{safe}}$ and following barrier certificate is satisfied.

\vspace{-0.35cm}
\begin{align}
\begin{gathered}
  \nabla {{h}}\left( \text{x} \right)f\left( x \right)+\text{ }\!\!~\!\!\text{ }\nabla {{h}}\left( \text{x} \right)\text{g}\left( \text{x} \right){{u}^{i+1}}+Z\,{{h}}\left( x \right)\\ -\nabla {{h}}\left( x \right)g(x){{(\nabla {{h}}\left( x \right)g(x))}^{T}}\,\,\text{is }\,\,\text{SOS }\!\!~\!\!\text{ }      \\ 
\end{gathered} 
\label{eq43}
\end{align}

\noindent
With an initial robust safe control policy ${{u}^{0}}$, the safety of the control policy can always be guaranteed in each iteration.

We now show globally robust stability of the control solution when there is no conflict between stability and safety. When there is no conflict, $\delta =0$ because in this case, no relaxation variable is needed to guarantee feasibility by selecting a suitable scalar ${{\text{k}}_{\delta }}$. Now, develop an induction as follows.

i) Suppose $i=1$. Under Assumption 4, ${{u}^{1}}$ is globally robust stabilizing, and we also have ${{V}^{1}}\in P$. For each ${{x}_{0}}\in D$ and ${{x}_{0}}\ne 0$, we can obtain

\vspace{-0.25cm}
\begin{equation}
{{V}^{1}}({{x}_{0}})\ge \int\limits_{0}^{\infty }{\overline{r}(x,{{u}^{1}})dt}>0
\label{eq44}
\end{equation}
\vspace{-0.2cm}

\noindent 
Using this inequality and the constraint in (\ref{eq35}), under Assumption 1 it follows that 

\vspace{-0.25cm}
\begin{equation}
{{V}^{o}}\le {{V}^{_{1}}}\le {{V}^{_{0}}}
\label{eq45}
\end{equation}
\vspace{-0.4cm}

\noindent 
since ${{V}^{o}}$ and ${{V}^{0}}$ are assumed to be positive definite, obviously ${{V}^{{}}}\in P$.

ii) Assume ${{u}^{i-1}}$ is globally robust stabilizing, and ${{V}^{i-1}}\in P$ for $i\ge 2$. We now prove that ${{u}^{_{i}}}$ is globally robust stabilizing and ${{V}^{i}}\in P$. Along the system trajectory of system (1) and $u={{u}^{_{i}}}$ the following holds                              

\vspace{-0.25cm}
\begin{equation}
{{\dot{V}}^{i-1}}={{(\nabla {{V}^{i-1}})}^{T}}(f+g({{u}^{i}}+\omega))=-L({{V}^{i-1}},{{u}^{i}})-\overline{r}(x,{{u}^{i}})\le 0
\label{eq46}
\end{equation}
\vspace{-0.4cm}

\noindent
Therefore, ${{u}^{i}}$ is globally robust stabilizing in this situation. ${{V}^{i-1}}$ is a Lyapunov function for the system and we have

\vspace{-0.25cm}
\begin{equation}
{{V}^{i}}({{x}_{0}})\ge \int\limits_{0}^{\infty }{\overline{r}(x,{{u}^{i}})dt},\,\,\,\,\,\forall {{x}_{0}}\ne 0.
\label{eq47}
\end{equation}
\vspace{-0.4cm}

\noindent
Similarly, the following inequalities hold
 
\vspace{-0.25cm}
\begin{equation}
{{V}^{o}}({{x}_{0}})\le {{V}^{_{i}}}({{x}_{0}})\le {{V}^{_{i-1}}}({{x}_{0}})
\label{eq48}
\end{equation}
\vspace{-0.4cm}

\noindent
Since ${{V}^{o}}$ and ${{V}^{_{i-1}}}$ are assumed to be positive definite, obviously ${{V}^{_{i}}}\in P$.

3) By 2), the sequence $\left\{ {{V}^{i}}(x) \right\}_{i=0}^{\infty }$ is monotonically decreasing with $0$ as their lower bound due to its positive definite property. Therefore, there exists a lower bound ${{V}^{*}}(x)$ such that $\underset{i\to \infty }{\mathop{\lim }}\,{{V}_{i}}(x)={{V}^{*}}(x)$. Let $\left\{ {{p}_{i}} \right\}_{i=0}^{\infty }$ be a sequence for $\left\{ {{V}^{i}}(x) \right\}_{i=0}^{\infty }$ such that ${{V}^{i}}=p_{i}^{T}{{\overrightarrow{m}}_{2,2r}}(x)$ so $\underset{i\to \infty }{\mathop{\lim }}\,{{p}_{i}}={{p}^{*}}\in {{\mathbb{R}}^{{{n}_{2r}}}}$, ${{V}^{*}}={{p}^{*}}{{\overrightarrow{m}}_{2,2r}}(x)$. Similarly, it can also be shown that ${{V}^{o}}(x)\le {{V}^{*}}(x)\le {{V}^{0}}$. Since ${{V}^{o}}$ and ${{V}^{_{i-1}}}$ are assumed to be positive definite, ${{V}^{*}}\in \mathbb{R}{{[x]}_{2,2r}}$ and is positive definite.

4) By 3), we know that  

\vspace{-0.25cm}
\begin{equation}
\overline{H}(V^{*})=-L(V^{*},u^{*})\le 0.
\label{eq50}
\end{equation}
\vspace{-0.4cm}

\noindent So, $V^{*}$ is a solution to Problem 1, the inequality in 4) can be derived from the fourth property of Theorem 5. This completes the proof.            $\hfill$ $\square$   

\vspace{-0.2cm}

\section{SIMULATION RESULTS }
In this section, the peoposed safe optimal control algorithm is applied to car suspension system. Consider a model of a car suspension system as                                        
\smallskip
\smallskip

\begin{align}
\begin{gathered}
\left[ {\begin{array}{*{20}{c}}
{{{\dot x}_1}}\\
{{{\dot x}_2}}\\
{{{\dot x}_3}}\\
{{{\dot x}_4}}
\end{array}} \right] = \left[ {\begin{array}{*{20}{c}}
{{x_2}}\\
{\frac{{{K_s}({x_3} - {x_1}) - {K_n}{{({x_1} - {x_3})}^3} + {B_s}({x_4} - {x_2}) + cu}}{{{M_b}}}}\\
{{x_3}}\\
{\frac{{{K_s}({x_3} - {x_1}) - {K_n}{{({x_1} - {x_3})}^3} + {B_s}({x_4} - {x_2}) + {K_1}{x_3} + cu}}{{{M_w}}}}
\end{array}} \right]
\end{gathered} 
\label{EXP3}
\end{align} 
\smallskip

\noindent
where ${{x}_{1}},{{x}_{2}},{{x}_{3}},{{x}_{4}}$ are position, velocity of the car body, position, velocity of the wheel assembly. ${{M}_{b}},\,{{M}_{w}}$ denote mass of car and wheel assembly respectively. ${{K}_{l}},{{K}_{s}},{{K}_{n}}$ are the tire stiffness, nonlinear suspension stiffness and linear suspension stiffness, respectively, ${{B}_{s}}$ is the damping rate of suspension and $c$ is a constant control signal related to the input force. In this experiment, the parameters of system are set to ${{M}_{b}}=300\,kg,\,{{M}_{w}}=60\,kg,\,{{B}_{s}}=1000\,Ns/m,\,{{K}_{s}}=16000\,N/m,\,{{K}_{t}}=190000\,N/m,\,{{K}_{n}}=1600\,N/m$. The safe region ${{\mathcal{X}}_{o}}$ is defined as                                           
\begin{align}
\begin{gathered}
\mathcal{X}_o =\{x|x\in {{\mathbb{R}}^{4}},\,-20 \leq x_4 \leq 25\}.
\end{gathered} 
\label{eq59}
\end{align}

The parameters in performance index are set to $q(x)=100x_1^2+x_2^2+x_3^2+x_4^2$ and $R=1$. We are interested in proving the system in the following set

\begin{equation}
\Theta =\{x|x\in {{\mathbb{R}}^{4}},\,\,\,|{{x}_{1}}|,\,|{{x}_{3}}|\le 0.5\,\,\,|{{x}_{2}}|,\,|{{x}_{4}}|\le 10\}.
\label{eq61}
\end{equation}

The state trajectories under the proposed safe optimal controller and uncontrolled condition are shown in Fig. \ref{f1} while the visualization of the value function, and safe optimal control signal are presented in Fig. \ref{f2} and Fig. \ref{f3}. According to Fig. \ref{f2}, the sequence of value functions evaluated by (\ref{eq35}) is monotonically decreasing, and it will reach a much smaller value than the initial one after 10 iterations. Besides, it can be clearly observed that the trajectory of $x_4$ under the proposed algorithm will not violate the safe constraints (\ref{eq59}).

\vspace{-0.2cm}
\begin{figure}[!ht]
\begin{center}
\includegraphics[width=0.9\columnwidth,height=95mm]{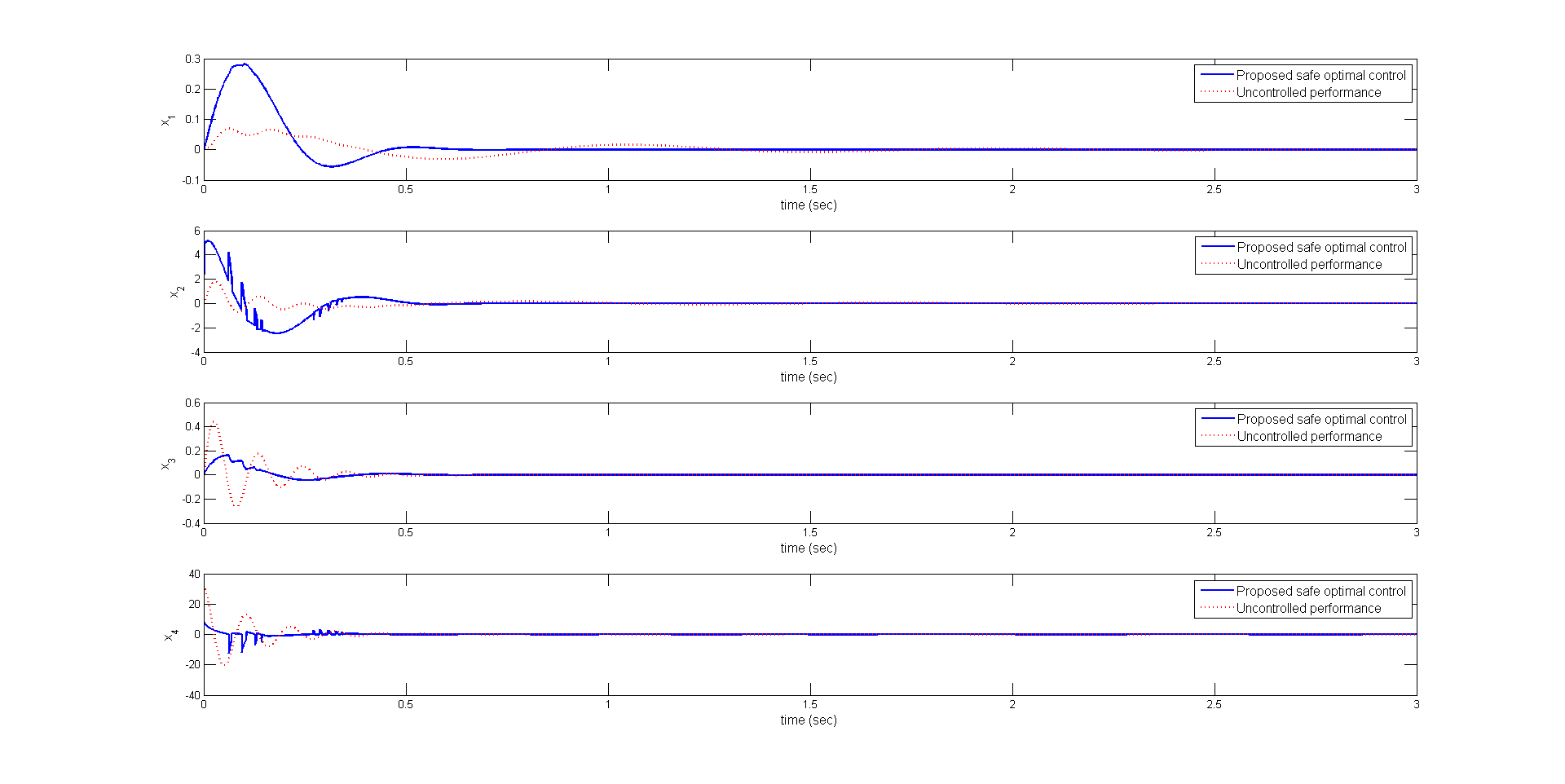}
\vspace{-5pt}\caption{The system trajectories of car suspension system under the proposed safe optimal control design.}
\label{f1}
\captionsetup{justification=centering}
\end{center}
\vspace{-0.2cm}
\end{figure}

\vspace{-0.2cm}
\begin{figure}[!ht]
\begin{center}
\includegraphics[width=0.7\columnwidth,height=60mm]{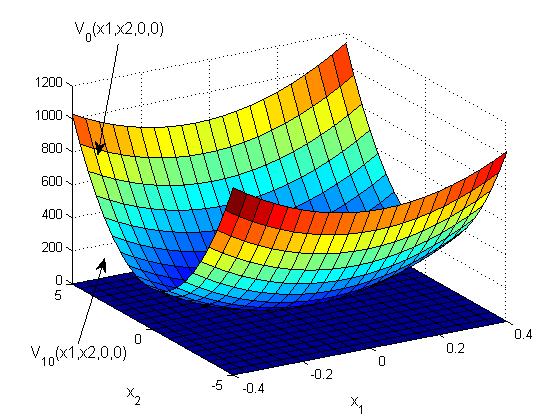}
\vspace{-5pt}\caption{The 3D plot of value function of car suspension system for the initial control policy at the first iteration and the final control policy found after iteration 10.}
\label{f2}
\captionsetup{justification=centering}
\end{center}
\vspace{-0.2cm}
\end{figure}

\vspace{-0.2cm}
\begin{figure}[!ht]
\begin{center}
\includegraphics[width=0.95\columnwidth,height=65mm]{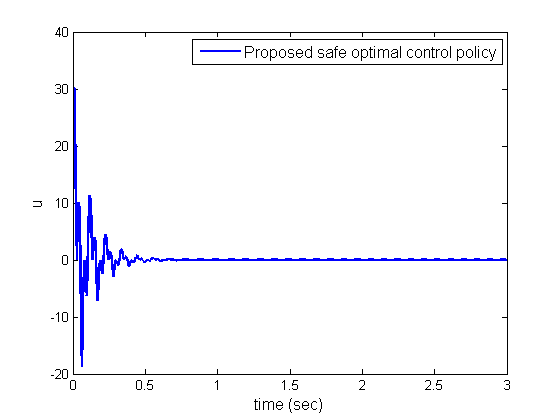}
\vspace{-5pt}\caption{The control signal for the proposed safe optimal control when applied to the suspension system.}
\label{f3}
\captionsetup{justification=centering}
\end{center}
\end{figure}

\smallskip
\smallskip
\smallskip
\smallskip
\smallskip
\smallskip

To analyze and verify the effectiveness of the proposed algorithm, in the following subsections, two comparison experiments are conducted. Since the optimal control of safety-critic systems is the concern of the paper, satisfaction of safe constraints (i.e., staying forever in the safe regions when starting from the safe set) as well as optimality of the proposed control policy are investigated  in the following subsections. 

\subsection{Safety Verification of the Proposed Algorithm}
We now compare our results with an optimal control algorithm presented in \cite{a28}. Implementing the algorithm in \cite{a28} to the car suspension system (\ref{EXP3}) results in the system performance shown in Fig. \ref{f4} and control signal shown in Fig. \ref{f5}. From Fig. \ref{f4}, it can be seen that the trajectory of $x_4$ controlled by \cite{a28} violates safety constraints (\ref{eq59}). Compared with \cite{a28}, the proposed safe optimal control policy will not escape the safe set (\ref{eq59}. Therefore, the proposed algorithm guarantees safety while the algorithm presented in \cite{a28} violates safety in this simulation case.

\vspace{-0.2cm}
\begin{figure}[!ht]
\begin{center}
\includegraphics[width=1.09\columnwidth,height=105mm]{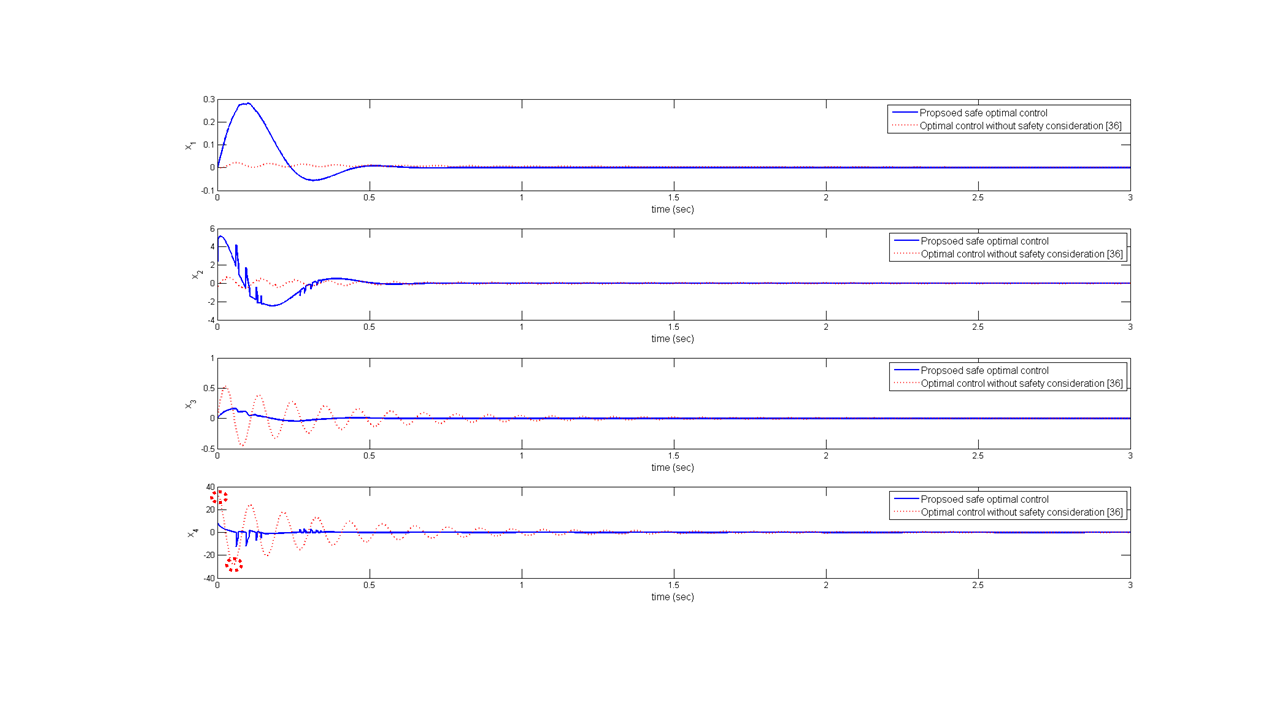}
\vspace{-25pt}\caption{Comparison of the trajectories of the suspension system under the proposed safe optimal control design and the optimal control design without safety guarantee presented in  \cite{a28}.}
\label{f4}
\captionsetup{justification=centering}
\end{center}
\vspace{-0.4cm}
\end{figure}

\vspace{-0.2cm}
\begin{figure}[!ht]
\begin{center}
\includegraphics[width=1.05\columnwidth,height=75mm]{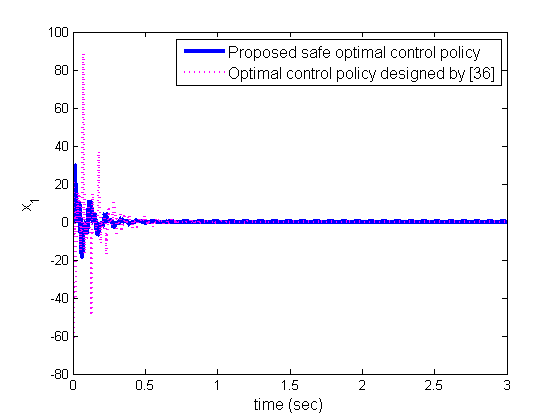}
\vspace{-5pt}\caption{Comparison of the control signal of the suspension system under the proposed safe optimal control design and the optimal control design without safety guarantee presented in  \cite{a28}.}
\label{f5}
\captionsetup{justification=centering}
\end{center}
\vspace{-0.2cm}
\end{figure}

\subsection{Optimality Verification of the Proposed Algorithm}

We now compare our proposed safe optimal control design method with the safe control design method presented in \cite{safe1}. The method in \cite{safe1} only considers safety and stability and does not incorporate any long-horizon optimality in the control design phase.  The simulation results for the suspension system for the controller desiged without considering any cost optimization in \cite{safe1} are shown in the Fig. \ref{f6}. It can be clearly observed from Fig. \ref{f6} that both design methods will not violate the safety constraints (\ref{eq59}) (represented by the blue dash line). However, as shown in Fig. \ref{f7}, the value function for corresponding to the proposed safe optimal controller is much smaller than the value function corresponding to the same reward function calculated for the safe controller. This clearly shows that our proposed approach outperforms the standard safe control design approaches. 

\vspace{-0.2cm}
\begin{figure}[!ht]
\begin{center}
\includegraphics[width=1.18\columnwidth,height=120mm]{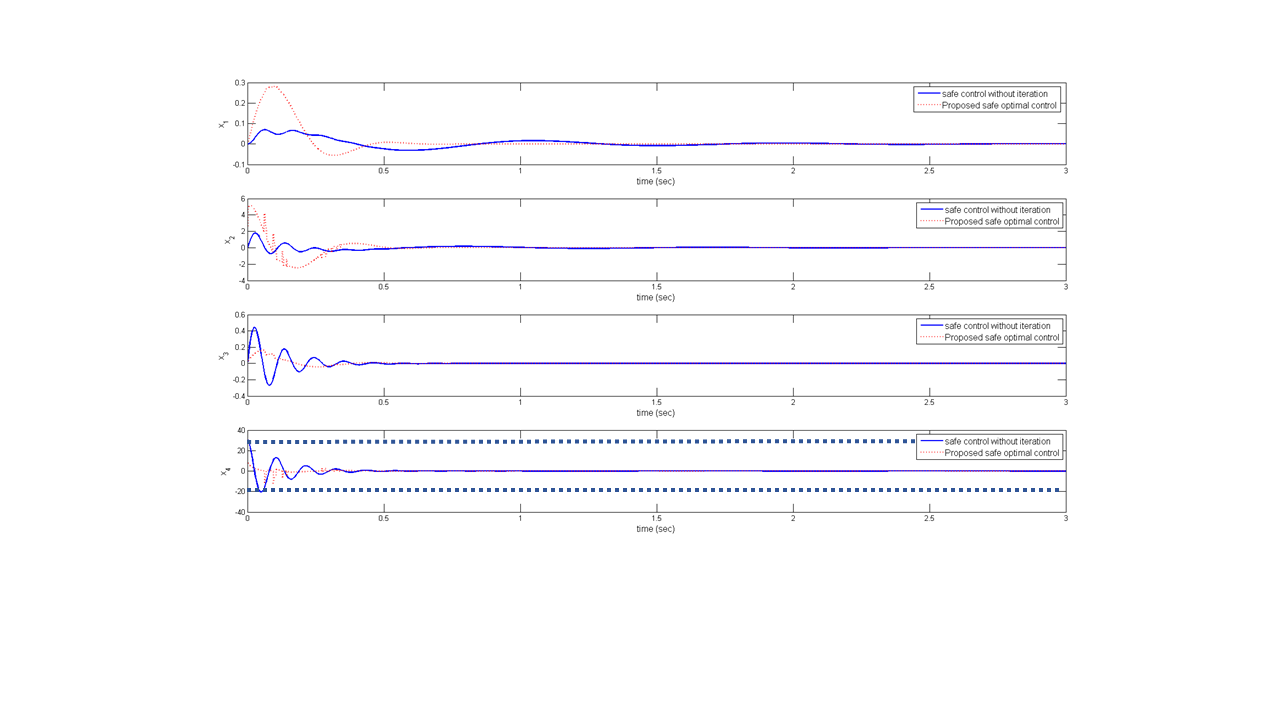}
\vspace{-75pt}\caption{Comparison of the trajectories of car suspension system under the proposed safe optimal control and  safe control design without optimality consideration.}
\label{f6}
\captionsetup{justification=centering}
\end{center}
\vspace{-0.3cm}
\end{figure}

\vspace{-0.2cm}
\begin{figure}[!ht]
\begin{center}
\includegraphics[width=1.18\columnwidth,height=85mm]{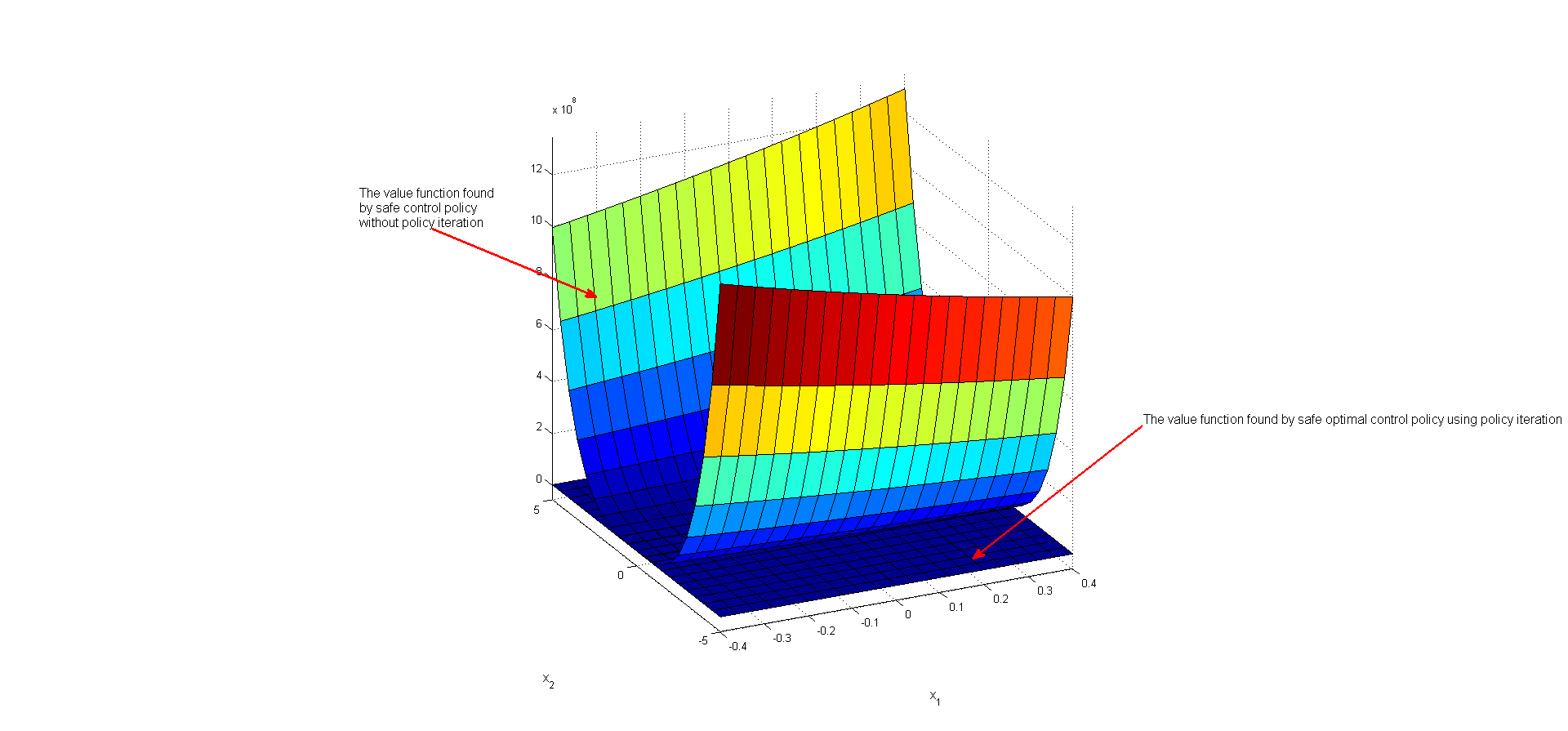}
\vspace{-5pt}\caption{Comparison of the value function of the system under the proposed safe optimal control and  safe control design without optimality consideration.}
\label{f7}
\captionsetup{justification=centering}
\end{center}
\vspace{-0.4cm}
\end{figure}

\section{CONCLUSION AND FUTURE WORK }

In this paper, the problem of both safe control and optimal control are investigated simultaneously. The optimal solution is derived by solving the Bellmen inequality using SOS. The obtained result is then verified by the barrier function to guarantee the safety of system. A relaxation variable is added to handle conflict between safety and stability of the system. The final controller obtained from the proposed method is not necessarily an optimal one but assures the safety of the system with guaranteed performance, called a satisficing solution. Numerical simulations are given to illustrate the effectiveness of the proposed algorithm. The proposed algorithm is applied to the car suspension system. Possible extensions of the presented work for the tracking problem, event-triggered systems, control of systems with unmatched disturbance and output regulation will be explored in the future.

\appendices
\section{PROOF OF THE THEOREM 4}

1) Define ${{u}_{0}}(x)=-\frac{1}{2}{{R}^{-1}}{{(\nabla V_{^{0}}^{T}\left( x \right)g(x))}^{T}}$, since (\ref{eq26}) holds, then

\begin{align}
\begin{gathered}
   \overline{H}({{V}_{0}})=\nabla V_{0}^{T}(f(x)+g(x){{u}_{0}}+w)+\overline{r}(x,{{u}_{0}}) \hfill\\  
 \quad \quad \quad =\nabla V_{0}^{T}(f(x)+g(x){{u}_{1}}+w)+\overline{r}(x,{{u}_{1}})+\nabla V_{0}^{T}g(x)({{u}_{0}}-{{u}_{1}}) \hfill\\  
 \quad \quad \quad +u_{0}^{T}R{{u}_{0}}-u_{1}^{T}R{{u}_{1}}+{{\beta }_{{{u}_{0}}}}(x)-{{\beta }_{{{u}_{1}}}}(x) \hfill\\  
 \quad \quad \quad =\nabla V_{0}^{T}(f(x)+g(x){{u}_{1}}+w)+\overline{r}(x,{{u}_{1}})-{{({{u}_{0}}-{{u}_{1}})}^{T}}R({{u}_{0}}-{{u}_{1}})+{{\beta }_{{{u}_{0}}}}(x)-{{\beta }_{{{u}_{1}}}}(x) \hfill\\  
 \quad \quad \quad \le 0 \hfill\\
  \end{gathered} 
\label{eq70}
\end{align}

\noindent
Therefore, ${{V}_{0}}$ is a feasible solution for (\ref{eq27}). Now, let us prove the solution to Problem 1 is unique.

2) Consider system (\ref{eq1}) and control policy (\ref{eq28}). Indeed, along the solutions of the closed-loop system, it follows that:     

\begin{align}
\begin{gathered}
   \dot{V}=\nabla {{V}^{T}}\left( x \right)\left( f\left( x \right)+g\left( x \right)\overline{u}+\omega(t) \right)\,  \hfill\\ 
  \quad  =\nabla {{V}^{T}}\left( x \right)f(x)+\nabla {{V}^{T}}\left( x \right)g(x)\overline{u}+\nabla {{V}^{T}}\left( x \right)\omega(t)  \hfill\\ 
 \end{gathered} 
\label{eq73}
\end{align}

\noindent
Since Problem 1 performs optimization over all value functions but satisfies $\overline{H}(V)\le 0$. Then applying Hamilton function $\overline{H}\text{(}V\text{)=}\nabla {{V}^{T}}(x)\,(f(x)+g(x)u)+\overline{r}\left( x,u \right)$ of (\ref{eq11}), one has

\begin{align}
\begin{gathered}
  \dot{V}\le -\overline{r}(x,\overline{u})+\nabla {{V}^{T}}\left( x \right)\omega(t)   \hfill\\ 
  \quad  =-q\left( x \right)-{{\overline{u}}^{T}}R\overline{u}-d_{max}^TRd_{max}-\beta_u(x)+\nabla {{V}^{T}}\left( x \right)g(x)d  \hfill\\ 
    \quad  =-2{{\overline{u}}^{T}}Rd-d_{max}^TRd_{max}-q(x)-{{\overline{u}}^{T}}R\overline{u}-\beta_u(x) \hfill\\
      \quad =-d_{max}^TRd_{max}-q(x)+{{d}^{T}}(x)Rd(x)-{{(d+\overline{u})}^{T}}R(d+\overline{u})-\beta_u(x)\hfill\\
  \quad \le -d_{max}^TRd_{max}-q(x)+{{d}^{T}}(x)Rd(x)-\beta_u(x) \hfill\\
   \quad \le -q(x)-\beta_u(x)\le -q(x)-\overline{u}_{{}}^{T}R{{\overline{u}}_{{}}}<0 \hfill\\
 \end{gathered} 
\label{eq74}
\end{align}

From (\ref{eq74}), it is easily to see $V$ is a well-defined Lyapunov function for closed-loop system. Therefore, $\overline{u}$ is global robust stable. when there is no conflict, select $\text{ }\!\!\delta\!\!\text{ =0}$ and $b(x)=1$, it is obvious that solution belong to (20) when ${{p}_{s}}(u,s)=-{{(\nabla V)}^{T}}\left( x \right)\left( f\left( x \right)+g\left( x \right){{u}_{0}} \right)$ and ${{p}_{r}}(u,s)=r(x,{{u}_{0}})$.

3) Next we show that index (\ref{eq4}) has an upper bound. By integrating both sides of inequality (\ref{eq74}) over time $[0,T]$, we derive:

\vspace{-0.4cm}
\begin{equation}
V({{x}_{0}})-V(x(T))\ge \int\limits_{0}^{T}{(q\left( x \right)+{{u}^{T}}Ru})dt
\label{eq75}
\end{equation}
\vspace{-0.2cm}

\noindent
Since $V$ is a well-defined Lyapunov function of system (\ref{eq1}), we have $V(x(T))\to 0$ as $T\to \infty $. Thus, (\ref{eq75}) yields

\vspace{-0.4cm}
\begin{equation}
V({{x}_{0}})\ge \int\limits_{0}^{\infty }{(q\left( x \right)+{{u}^{T}}Ru})dt
\label{eq76}
\end{equation}
\vspace{-0.2cm}

\noindent
which implies that $J\left( {{x}_{0}},u \right)\le V({{x}_{0}})$ for all small bounded disturbance $d$. This shows performance index (\ref{eq4}) has an upper bound $V({{x}_{0}})$. And along the trajectory of the nominal system $\dot{x}=f\left( x \right)+g\left( x \right)u$, we have  

\vspace{-0.4cm}
\begin{equation}
\dot{V}=\nabla {{V}^{T}}\left( x \right)\left( f\left( x \right)+g\left( x \right)u \right)\le -q\left( x \right)-{{u}^{T}}Ru-\beta_u(x)-d_{max}^TRd_{max}
\label{eq77}
\end{equation}
\vspace{-0.2cm}

\noindent
Integrating both sides of (\ref{eq77}) over $[0,T]$ yields

\vspace{-0.4cm}
\begin{equation}
V({{x}_{0}})-V(x(T))\ge \int\limits_{0}^{T}{(q\left( x \right)+{{u}^{T}}Ru}+\beta_u(x)+d_{max}^TRd_{max})dt
\label{eq78}
\end{equation}
\vspace{-0.2cm}

\noindent
Letting $T\to \infty$, we obtain $\overline{J}\left( {{x}_{0}},u \right)\le V({{x}_{0}})$.

4) By 3), we know

\vspace{-0.4cm}
\begin{equation}
V({{x}_{0}})\ge \bar{J}({{x}_{0}},\overline{u})\ge \underset{u}{\mathop{\min }}\,\bar{J}({{x}_{0}},\overline{u})={{V}^{o}}({{x}_{0}})
\label{eq79}
\end{equation}
\vspace{-0.2cm}

\noindent
Therefore, the second inequality in (\ref{eq29}) is proved. Besides,

\begin{align}
\begin{gathered}
  \overline{H}(V)=\overline{H}(V)-\overline{H}({{V}^{o}})   \hfill\\ 
    =-{{(\nabla V-\nabla {{V}^{o}})}^{T}}(f+g\overline{u})+\overline{r}(x,\overline{u})-{{(\nabla {{V}^{o}})}^{T}}g({{u}^{o}}-\overline{u})-\overline{r}(x,{{u}^{o}})   \hfill\\ 
  ={{(\nabla V-\nabla {{V}^{o}})}^{T}}(f+g\overline{u})+{{(\overline{u}-{{u}^{o}})}^{T}}R(\overline{u}-{{u}^{o}})+\beta_{\bar{u}}(x)-\beta_{u^o}(x)  \hfill\\ 
    \ge {{(\nabla V-\nabla {{V}^{o}})}^{T}}(f+g\overline{u})  \hfill\\ 
 \end{gathered} 
\label{eq80}
\end{align}

\noindent
Integrating above equation along the solutions of the closed-loop system (\ref{eq1}) with control policy (\ref{eq28}) on the interval $[0.\infty]$, we derive

\vspace{-0.4cm}
\begin{equation}
V({{x}_{0}})-{{V}^{o}}({{x}_{0}})\le -\int\limits_{0}^{\infty }{\bar{H}(V(x(t)))dt}
\label{eq81}
\end{equation}
\vspace{-0.2cm}

5)  By 3), for any feasible solution $V$ to  (\ref{eq27}), we have $V(x)\ge {{V}^{o}}(x)$. Therefore

\vspace{-0.4cm}
\begin{equation}
\mathop{\int }_{\Omega }^{{}}{{V}^{o}}\left( x \right)dx\le \mathop{\int }_{\Omega }^{{}}V\left( x \right)dx
\label{eq82}
\end{equation}
\vspace{-0.2cm}

\noindent
which concludes ${{V}^{o}}$ is a global robust optimal solution to (\ref{eq27}).

The proof is complete.            $\hfill$ $\square$

\end{document}